# Accelerating quantum materials development with advances in transmission electron microscopy

*Parivash Moradifar[1], Yin Liu[1,2], Jiaojian Shi[1,3], Matti Lawton Siukola Thurston[1], Hendrik Utzat[1,4], Tim B. van Driel[5], Aaron M. Lindenberg[1,3], Jennifer A. Dionne[1,6]*

[1]Department of Materials Science and Engineering, Stanford University, Stanford, California 94305, USA

[2]Department of Materials Science and Engineering, North Carolina State University, Raleigh, North Carolina 27695, USA

[3]SLAC National Accelerator Laboratory, 2575 Sand Hill Road MS69, Menlo Park, California 94025, USA

[4]Department of Chemistry, University of California Berkeley, Berkeley, California 94720, USA

[5]Linac Coherent Light Source, SLAC National Accelerator Laboratory, 2575 Sand Hill Road, Menlo Park, California 94025, USA

[6]Department of Radiology, Stanford University, Stanford, California 94305, USA

## Abstract

Quantum materials are driving a technology revolution in sensing, communication, and computing, while simultaneously testing many core theories of the past century. Materials such as topological insulators, complex oxides, superconductors, quantum dots, color center-hosting semiconductors, and other types of strongly correlated materials can exhibit exotic properties such as edge conductivity, multiferroicity, magnetoresistance, superconductivity, single photon emission, and optical-spin locking. These emergent properties arise and depend strongly on the material's detailed atomic-scale structure, including atomic defects, dopants, and lattice stacking. In this review, after the introduction of different classes of quantum materials and quantum excitations, we describe how progress in the field of electron microscopy (EM), including in-situ and in-operando-EM, can accelerate advances in quantum materials. Our review describes EM methods including: i) principles and operation modes of EM; ii) EM spectroscopies, such as electron energy loss spectroscopy (EELS), cathodoluminescence (CL), and electron energy gain spectroscopy (EEGS); iii) four-dimensional scanning transmission electron microscopy (4D-STEM); iv) dynamic and ultrafast EM (UEM); v) complimentary ultrafast spectroscopies (UED, XFEL); and vi) atomic electron tomography (AET). We discuss how these methods inform structure-function relations in quantum materials down to the picometer scale and femtosecond time resolution, and how they enable precision positioning of atomic defects and high-resolution manipulation of quantum materials. Among numerous notable results, our review highlights how EM has enabled identification of the 3D structure of quantum defects; measuring reversible and metastable dynamics of quantum excitations; mapping exciton states and single photon emission; measuring nanoscale thermal transport and coupled excitation dynamics; and measuring the internal electric field and charge density distribution of quantum heterointerfaces- all at the quantum materials' intrinsic atomic and near atomic-length scale. We conclude by describing open challenges for the future, combining ultralow temperature (below 10K) with atomic spatial resolution and meV energy resolution. With atomic manipulation and ultrafast characterization enabled by EM, quantum materials will be poised to integrate into many of the sustainable and energy-efficient technologies needed for the 21st century.

## 1. Introduction

The Solvay Conference of 1927, aptly titled "Electrons and Photons", gathered world-renowned scientists to discuss the newly formulated theories of quantum mechanics. Nearly 100 years later, armed with materials growth techniques that enable near-atomic-scale structuring of materials, scientists are still discovering novel, emergent properties linked to quantum mechanical behavior.[1] These quantum properties not only defy our intuition about the physical world but also hold promise for improved and energy-efficient sensing, imaging, communications, and computing systems.[2]

While all matter is governed by quantum mechanical behavior, 'quantum materials' can be defined as materials which exhibit exotic properties arising from the non-classical nature of constituent particles.[3] Examples of such properties include, but are not limited to, electrical conductivity (high-temperature superconductivity, topological superconductivity), magnetism (multiferroicity, colossal magnetoresistivity), and quantum-optical behavior (single-photon emission, quantized nonlinear optics, and optical spin-locking).[4–8] These properties are governed by quantum excitations and quasi-particles such as electrons, magnons, excitons, and polaritons, among others, which can further feature strong correlations and topological effects. Probing and tuning the characteristics of these quasiparticles is important for the manipulation of a material's macroscopic properties.

Figure 1 highlights examples of quantum materials, their elementary quantum excitations, and their relevant energy ranges. As seen, quantum materials systems include helimagnetic materials, complex oxides/cuprates, quasicrystals, low-dimensional materials, topological insulators, and metamaterials, with excitations spanning THz to ultraviolet frequencies. These material classes host magnons, charge density waves, phonons, polaritons, excitons, and Mott Hubbard excitations, which are responsible for their exotic functionalities. For example, magnetic skyrmion materials host peculiar fingerprints associated with their spin configurations at terahertz frequencies; these can be used for high speed information processing, high density and ultra-low energy consumption digital storage.[9,10] In parallel, excitons as correlated electron-hole pairs are promising candidates as data carriers for ultrafast and robust information transfer, replacing free electrons.[11] In addition, low-dimensional materials and quantum dots have reduced dielectric screening that gives rise to unique excitonic properties, with no direct counterpart in the material's bulk form; these properties include tunable bandgaps, strong spin-orbit coupling, and Stark-tunability,[12–14] and can form the basis for optically-addressable, deterministically-positioned, and reconfigurable spin-qubits. Such a rich palette of quantum materials emphasizes the enormous opportunities in this booming field and serves as a general guideline for our review.

One example of a particularly exciting emerging class of quantum materials is two-dimensional materials and their heterostructures, with quantum excitations in the near-infrared (near-IR) and visible (Vis) frequency ranges but also including vibrational excitations down to THz frequencies.[15–17] These atomically thin, van der Waals (vdW) materials can be fabricated over large areas and stacked with atomic precision to form heterostructures. Depending on the stacking angle, these heterostructures can form moiré patterns that induce entirely new band structures and potential wells for electrons, holes, and excitons.[18,19] In moiré metamaterials, strong interlayer electronic interactions exhibit completely different properties than their constituent layers, including topological protection,[20,21] positionally-precise single-photon emission,[22–27] and even the possibility of superconductivity.[28,29] Two-dimensional materials and their heterostructures can be interfaced with other photonic, plasmonic, and metasurface cavities to modify the local density of optical states,[30] control radiative quantum efficiency,[31] achieve directional emission,[32] enhance coherence times,[33,34] and modify coupling to plasmonic or magnetic degrees of freedom.[35–37]

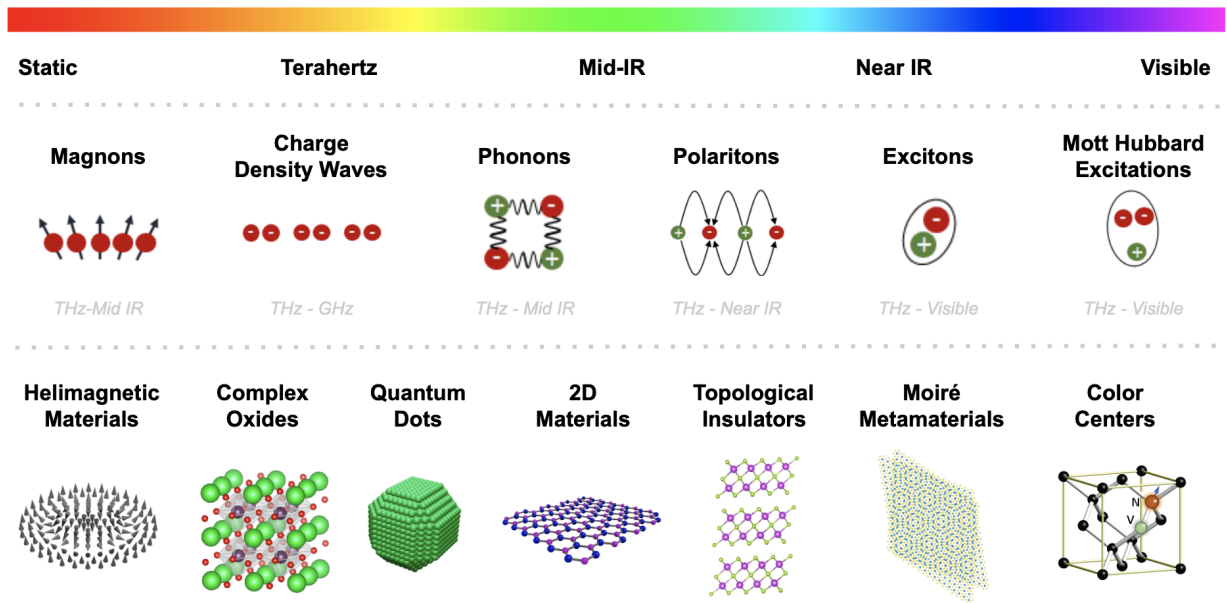

Figure 1. Quantum excitations covering a broad spectral range in various quantum platforms. The elementary excitations (magnons to Mott Hubbard excitations) and associated quantum material classes (magnetic materials to metamaterials and color centers) are arranged covering a broad frequency range (static, to visible).

Structural heterogeneities of quantum materials play an extremely important role in determining their emergent quantum properties. Such structural heterogeneities include, but are not limited to, defects (vacancies, antisites, dislocations etc.), defect-induced strain,[38] grain boundaries, heterointerfaces, impurity dopants, and edge configurations (zigzag, armchair, etc.). These atomic-structure variations can introduce scattering pathways (point defects),[39] limit carrier mobility, and create trapped states (grain boundaries and heterointerfaces)[40,41] as well as vary the bonding configuration and local charge distribution (impurity dopants).[42–45] However, defects and imperfections are not always detrimental.[46–48] For example, lattice disorder resulting from impurities and defects can lead to electron localization, known as Anderson localization, which is a crucial mechanism for metal-to-insulator transitions in condensed matter.[49,50] Additionally, atomic-scale defects can also be bright sources of single photons. For example, the negatively-charged nitrogen vacancy center (NV⁻) in diamond can serve as a qubit with long coherence times or as a single-photon quantum emitter.[51] The single-spin qubits associated with local NV centers further enable nanoscale magnetometry with quantum-limit noise.[52,53] From processing and synthesis aspect, there has been tremendous effort in engineering nanoscale heterogeneities through various deposition methods (CVD, ALD, MBE) and colloidal synthesis techniques, offering near-atomic precision manufacturing pathways.[54–58] Various external stimuli like high-energy plasma, ion/electron beam irradiation, quenching, and annealing[50] can also offer deterministic positioning of such heterogeneities.

To unlock the full potential of quantum materials, nanoscale imaging and analytical tools are required to probe the quantum properties at relevant length, time, and energy scales. This review covers recent progress in electron microscopy (EM) for probing quantum materials. For the last several decades, electron probes have been extensively used to acquire nanoscale insights, enabling new discoveries and contributing to the development and design of new materials in numerous cross-cutting disciplines. Today, modern EM offers significantly enhanced capabilities to probe materials in real and reciprocal space as well as in the energy, momentum, and time domains, with sub-angstrom spatial resolution, mili-electronvolt energy resolution and picosecond temporal resolution. Our review covers the basics of EM imaging and spectroscopy capabilities for quantum materials (section 2), as well as correlative electron microscopy and optically coupled EM capabilities with a focus on electron energy loss/gain spectroscopy (EELS, EEGS) and cathodoluminescence spectroscopy (CL) (section 3). We then discuss four-dimensional scanning transmission electron microscopy (4D-STEM) (section 4). We also cover dynamic and ultrafast TEM (UTEM) and compare with non-TEM based ultrafast techniques including X-ray free electron lasers (XFEL), ultrafast electron diffraction (UED) and ultrafast optical spectroscopies (section 5 and 6). We finally discuss atomic electron tomography (AET) (section 7), and cryogenic TEM at extremely low temperatures (future and outlook in section 8). We note that cryo-EM has been at the heart of many recent discoveries and several outstanding review papers covering this technique including opportunities and challenges.[59,60] Our manuscript concludes by discussing how future instrumentation advances can extend atomic-resolution capabilities to ultralow temperature, to directly visualize a myriad range of exotic quantum and electronic phenomena ranging from superconductivity to metal-insulator transitions.

## 2. Fundamental principles of transmission electron microscopy

Nobel Laureate Dr. Alan Finkel recently said, "without microscopy, there is no modern science." The earliest electron microscope was demonstrated by Max Knoll and Ernst Ruska in 1931, with the technique maturing over the past 80 years.[61] There are two major imaging modalities in transmission electron microscopy: conventional transmission electron microscopy (TEM) and scanning transmission electron microscopy (STEM). Under parallel illumination in TEM, high-energy electrons are transmitted through an electron-transparent specimen to form an image with magnifications up to a few hundred million times.[62] The operating accelerating voltages of TEM that define the energy of the incident electrons are typically between 60-300 kV with subsequent DeBroglie wavelength of incident electrons ranging from ~4.86-1.96 pm, respectively. The higher the energy of the electrons, the lower the wavelength, and thus the higher the resolution. STEM, as a derivative of TEM, is a powerful and versatile imaging modality applicable across a breadth of materials systems including quantum materials. In STEM, the electrons can focus to form a sub-angstrom (sub-Å) convergent probe with high probe current, allowing for operation modes ranging from imaging (real space) and diffraction (reciprocal space) to various spectroscopic analyses of materials (energy and momentum space). The size of the formed probe defines the ultimate spatial resolution, and the probe size can vary from a few nm down in conventional STEM to sub-atomic dimensions for an aberration corrected probe microscope.[63–65]

The aberrations and distortions of lenses are the major limiting factor for the ultimate spatial resolution in an EM. Aberrations can be caused by the energy difference of incident electrons, due to inhomogeneities/geometric aberrations of the lenses, or even contamination of lenses and apertures. Chromatic aberration can occur due to the energy difference of the electrons, while spherical aberration can occur due to the curvature and distortions of the field. Advances in aberration corrected electron microscopy over the last two decades have been achieved by compensating the spherical aberration ($C_s$) of round lenses using sets of multipole lenses generating negative $C_s$.[66–72] Aberration-corrected high-resolution transmission electron microscopy (AC-HRTEM) is capable of direct visualization of atomic structures with sub-Å resolution to resolve individual atoms and defects. Similarly, aberration-corrected high-resolution STEM (AC-HRSTEM), with the ability to identify easily interpretable atomic structures, heterogeneities, and spatially correlated electronic and photonic properties down to a single-atom level has been at the heart of unraveling new physical phenomena.[63,64]

Additionally, as the swift electrons interact with the specimen, it can initiate several processes including generating X-ray photons, optical photons, and inelastically scattered electrons. These signals can be measured using various electron-based spectroscopy and imaging techniques including X-ray energy dispersive spectroscopy (XEDS), cathodoluminescence (CL), electron energy loss spectroscopy (EELS) and electron diffraction (ED). Several comprehensive review papers have been recently published on various imaging [64,73] and analytical capabilities of TEM including EELS and XEDS.[74–76] A combination of these techniques, such as EELS/CL, XEDS and quantitative S/TEM provide insight on spatially correlating the microstructure, functional properties (electronic and optical properties), and chemical composition down to single-atom precision.

Figure 2 summarizes various steady-state and dynamic capabilities of an EM. Figure 2 (left panel) shows a schematic of an aberration-corrected STEM with implemented time-resolved capabilities. As seen, the electron beam is focused into a fine, intense, and aberration corrected probe at the specimen plane. Major components of an aberration corrected STEM include i) an electron gun that emits the high energy electrons (thermionic, Schottky or field emission), ii) an electron optical lens system, iii) aberration correctors (probe corrector in this schematic) and iv) detectors. The accelerated electron beam follows a helical trajectory parallel to Helmholtz coils in the column. The lens system includes condenser lenses, objective and projector lenses (post specimen lenses). The lens system is responsible for i) the formation and illumination of the beam (condenser lenses), ii) the imaging and initial magnification of the image (objective lenses) and iii) altering magnifications (projector lenses).[77–80]

In addition to the existing capabilities for probing in real, reciprocal, energy and momentum space, EM capabilities can be extended into the temporal domain. Extending EM capabilities into the temporal domain requires integration of pump-probe configurations based on ultrafast pulsed lasers into the electron microscopes to trigger electron emission from the gun at particular time intervals [80] as shown in the Figure 2 (left panel). A summary of dynamic and ultrafast TEM technique as well as representative examples are included in section 5. Ultrafast EM capabilities allow for the detection of electronic and atomic motions at their natural time and length scales.

In addition to time-resolved capabilities, Figure 2(right panel), highlights other in-situ/operando capabilities that can be combined in the sample region to study dynamic responses to various external stimuli. Such in-situ capabilities include introduction of reactive chemical environments, such as gas species [81–83] and liquids,[84,85] as well as other external stimuli like heat, cryo-temperatures, electrical biasing, optical biasing, and mechanical force. The extension of TEM capabilities into in-situ/operando imaging and spectroscopy has enabled real-time observation and monitoring of the structural evolution of materials, including sublimation, formation, and growth of nanoscale defects;[86–90] nanoscale switching phenomena such as ferroelectric and resistive switching;[91] phase transitions;[92] light-driven chemical transformations;[93] electrochemical reactions in lithium-ion batteries;[94,95] the electric-field response of domain walls;[96] nanoscale mapping of the charge recombination; and electrical triggered structural transformations such as nucleation of polarized domains.[97] Several reviews cover the basics, challenges, progress, and opportunities that each of these in-situ techniques can offer.[98–100]

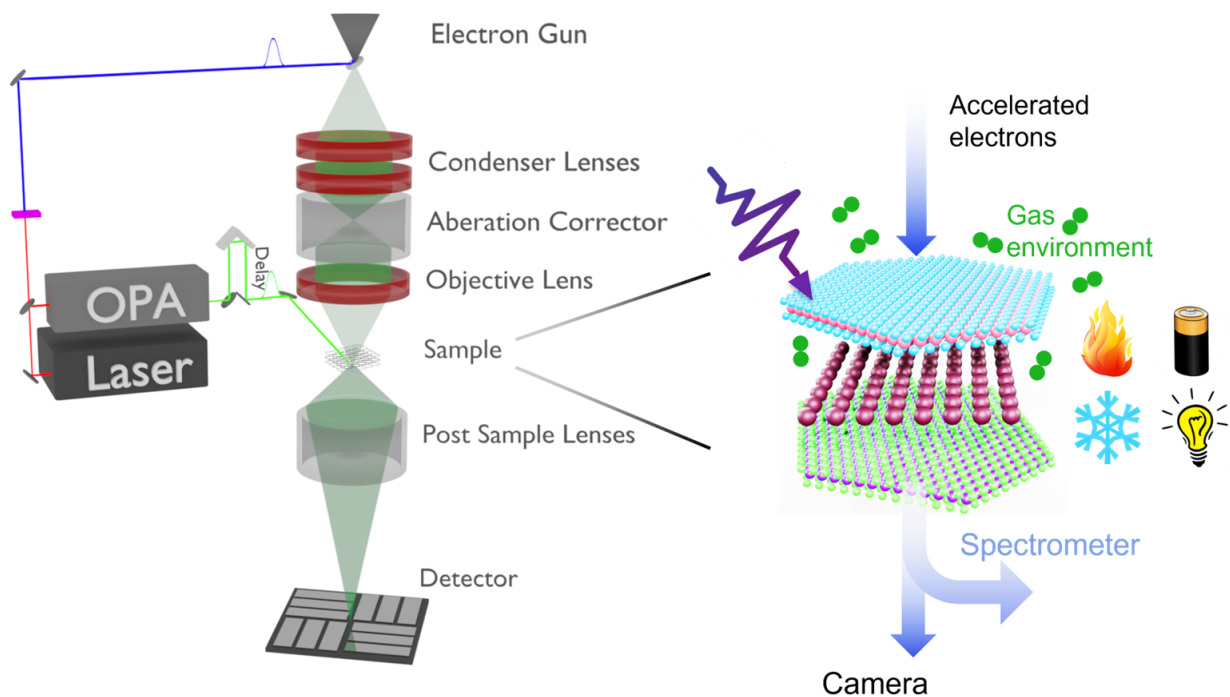

Figure 2. (left) Transmission electron microscopy and spectroscopy of quantum materials with ultrahigh spatial resolution as well as high temporal resolution when combined with an optical laser as illustrated. (right) Schematic of various in-situ EM capabilities showcasing the sample environment under various external stimuli.

## 3. Probing collective excitations in quantum materials using correlated electron and optical spectroscopy

Electron spectroscopies in EM include electron energy loss spectroscopy (EELS), electron energy gain spectroscopy (EEGS), and cathodoluminescence (CL) spectroscopy. These techniques probe a wide range of materials excitations with near atomic spatial resolution and meV energy resolution.[101–105] As the incident probe interacts with the material, the electron beam excites elementary quantum excitations, such as phonons, plasmons, excitons as well as inter- and intra-band transitions. In EELS, these excitations in turn lead to the loss of electron energy which can be measured and quantified.[75,106–108]

An electron energy loss (EEL) spectrum is the distribution of electron counts as a function of the energy-loss and is recorded by energy-dispersing the transmitted electrons via a magnetic prism. In EELS, the spectral energy resolution is determined by the spread of the primary electron energies emitted from the gun. A recorded EELS spectrum is classified in three regions with a decaying background: i) The zero loss peak (ZLP), ii) the valence or low-loss region, and iii) the core-loss region. The low-loss region and core-loss region are separated by an arbitrary boundary at ~50 eV. [75,109–113]

The ZLP is the feature with the highest intensity and represents elastic or quasi-elastic scattering events such as phonon excitations (with an energy below 100 meV). The low-loss regime represents inelastic scattering events such as collective excitations (excitonic states, surface plasmon resonances), and the inter- and intra-band transitions provide information on local electronic and optical properties. The core-loss region covers inelastic scattering events originating from inner-shell excitations containing information on local chemical properties including chemical concentration, oxidation, and bonding states. Furthermore, spectrum images (SI) containing both spatial and spectral information can be acquired by scanning a focused STEM probe over the sample (x,y probe position) and recording the corresponding EELS spectrum at each pixel.[75,109–113] Over the last decade, EELS has been shown to be a leading technique for probing elementary quantum excitations at their relevant length scales. Mapping polaritons, phonons, excitons and magnons is now possible using EELS with an energy resolution of a few meV. In addition, momentum-resolved EELS allows for the measurement of the electron energy loss as a function of the transferred momentum of the incident electrons. This enables the study of the dispersion and band structure covering the first Brillouin zone.[114–116]

Complementary to EELS, CL and EEGS spectroscopies have unique advantages in probing the optoelectronic properties of materials, such as excitonic properties in various materials systems including 2D semiconductors and defect emitters for quantum optics. In CL, one collects the electron-beam induced emission in the far-field to gain insight into the radiative decay pathways. CL spectroscopy in STEM, is a viable technique to distinguish radiative from non-radiative excitations in order to further identify and understand the nature of many elementary quantum excitations.[117–121] Electrons can also gain energy from the elementary excitations, a process utilized in EEGS. An incident light source on the material is used to pump excitations, essentially making EEGS the reverse of CL.[102,122] Similar to EELS, in CL and EEGS, the focused electron beam can be raster-scanned over the specimen to form a 2D spectral map. Therefore, these methods are capable of mapping structural and spectral information with ultrahigh precision.

Further insights can be derived by using CL, EEGS, and EELS in conjunction with the high spatial resolution of S/TEM imaging. Therefore, correlated electron and optical spectroscopy is a powerful experimental probe for revealing fundamental quasi-particle physics, including spin, charge, and lattice collective modes in quantum materials. Furthermore, these techniques can be paired with 4D-STEM capabilities and chemical composition analysis to enable a direct structure-property relationship.

## 3.1. Exploring terahertz-frequency excitations in van der Waals materials using momentum resolved EELS

Probing elementary excitations, such as spin waves and charge ordering states, requires meV energy resolution (see Figure 1). Momentum‐resolved EELS (M-EELS) also referred as q-EELS with ultrahigh energy resolution (a few meV), is a powerful technique for probing many such low energy elementary excitations.[123–128] Momentum‐resolved EELS has been used to reveal the Bose condensation of excitons in semimetal van der Waals materials.[127] A low-energy collective electronic mode-a plasmon mode around 80 meV- has been detected in the semimetal 1$T$-TiSe$_2$. Figure 3a highlights the momentum dependence of the valence plasmon in TiSe$_2$ for different
temperatures (300 K to 17 K). The dispersion curves along the (1,0) momentum direction (Figure 3b) are determined by fitting raw EELS spectra (Figure 3a). Figure 3b further highlights that the energy of the plasmon mode falls to zero at nonzero momentum around the phase-transition temperature at 190 K, indicating the dynamical slowing of plasma fluctuations. The M-EELS measurements provide evidence for exciton condensation at high temperature in TiSe$_2$ in which the plasmon mode behaves like the soft mode of a phase transition, demonstrating the condensation of electron-hole pairs at $T_C$.

In parallel, monochromated EELS using high energy electrons in TEM has been developed for studying low-energy excitations in condensed materials, with an improved energy resolution over the past two decades. An energy resolution of ~100 meV was first realized in monochromated STEM-EELS using a primary beam energy of 60-200 keV, which is sufficient to resolve optical excitations in the visible and near-infrared regime. Monochromated EELS has been extensively used for studying surface plasmon polaritons, optical resonant modes, excitons, and interband electronic transitions.[101,129,130] Recent progress in the use of monochromated EELS includes the use of novel monochromators, ultrahigh resolution electron analyzers, and high stability electronics, which in turn have enabled a state-of-the-art energy resolution of a few meV. In parallel, the capability for TEM imaging and spectroscopic analysis with nanoscale to atomic scale spatial resolution has been retained.[76,131–135]

Monochromated EELS (mono-EELS) has been used to directly characterize phonon polaritons (hybrid modes of lattice vibrations), and the local density of optical states in van der Waals 2D materials.[136,137] The phonon polaritons in van-der Waals materials are highly confined propagating modes with energies in the tens to hundreds of meV range. With decreasing thickness of van der Waals (vdW) materials approaching the single monolayer regime, the propagational wave vectors of polaritons significantly increase up to hundreds of times as compared to the vacuum wavelength. The detection of such polaritons requires techniques with extremely high spatial resolution that can not be obtained with most near-field optical scanning probe techniques. Using a focused electron beam, phonon-polaritons can be excited and detected in hBN thin films with thickness down to a monolayer.[136] Through 1D EELS mapping, the polaritons are visualized in real space with sub-nanometer spatial resolution, and with meV energy resolution at the sample/vacuum interface. Full dispersion mapping and quantification of the group velocities of phonon-polaritons in monolayer hBN has been demonstrated using mono-EELS with ultrahigh spatial and spectral resolution.

S/TEM-EELS with meV energy resolution and atomic scale spatial resolution has been further combined with momentum (angle) resolved capabilities.[116,138,139] The direction and magnitude of momentum transfer for an EELS measurement can be controlled through the azimuthal tilting angle of the incident electron beam with respect to the forward scattered beam. The transferred momentum is sufficiently large to cover the whole first Brillouin zone, and the momentum accuracy can be better than one percent of the Brillouin zone. There is nevertheless a tradeoff between the momentum and spatial resolution that can be simultaneously obtained, both of which are determined by the convergence angle of the electron beam. In previous studies, a converged beam excitation leads to higher spatial resolution (<2 nm) and a low momentum resolution (0.5 Å$^{-1}$), whereas parallel beam excitation leads to low spatial resolution (~40 nm) and a high momentum resolution (0.1 Å$^{-1}$). It is therefore important to identify the optimal experimental conditions to meet the spatial and momentum resolution requirements for a specific study.[116]

Momentum-resolved EELS has been used to measure the dispersion relation of various branches of bulk acoustic phonons and optical phonons in free-standing 2D materials including hBN and graphene with thickness down to a monolayer[139] as shown in Figure 3c. Direct nanoscale mapping of momentum-resolved excitations is particularly useful for spatially correlating the localization of quantum excitations, and can be used to unravel bulk, edge, and surface vibrational modes. A recent study has implemented momentum-resolved EELS mapping for studying the effect of crystalline defects on phonon propagation down to the single-defect level. In another study, a substantial intensity and energy change of acoustic phonons confined within a few nanometers around a single stacking fault in silicon carbide (SiC) has been detected in contrast to the defect-free materials which reveals phonon modes localized by a single planar crystal defect as shown in Figure 3d-f.[138]

Combining the ultrahigh energy and spatial resolution with momentum resolved capabilities, the S/TEM-EELS technique shows great promise to serve as one of the key techniques in quantum metrology frameworks. For instance, direct momentum-resolved EELS imaging is uniquely suited for understanding the critical impacts of crystalline heterogeneities including interfaces, atomic defects and local strains, as well as nanoscale charge, spin, and lattice order, all affecting the macroscopic properties of quantum materials. Furthermore, momentum-resolved low-loss EELS imaging can be combined with other TEM imaging, diffraction and spectroscopy capabilities such as 4D-STEM, EDS, and core-loss EELS to correlate the properties of low-energy excitations with the local structure and chemical environment down to the atomic level. To date, S/TEM-based q-EELS has demonstrated great success in revealing the vibrational modes in materials, which in turn can be applied to reveal the electron-lattice coupling and its effect on various electronic phase transitions in quantum materials. Beyond the study of vibrational lattice modes, the technique can be extended to measuring a wide range of low-energy excitations and various quantum phase transitions in emergent quantum systems, including polaritons, excitons and their Bose condensations, mid- and far-infrared surface plasmon in 2D materials, terahertz electronic states and plasmons, superfluidity, electronic and lattice ordering of charge density waves, spin waves in magnetic materials, Luttinger liquids, and novel quasiparticles such as polarons in high temperature superconductors. With further improvement in energy and momentum resolution, the S/TEM based q-EELS technique is expected to become an essential wave-vector-resolved probe alongside with angle resolved photoemission spectroscopy (ARPES), neutron scattering, inelastic X-ray scattering, and scanning tunneling microscopy for the study of quantum materials.

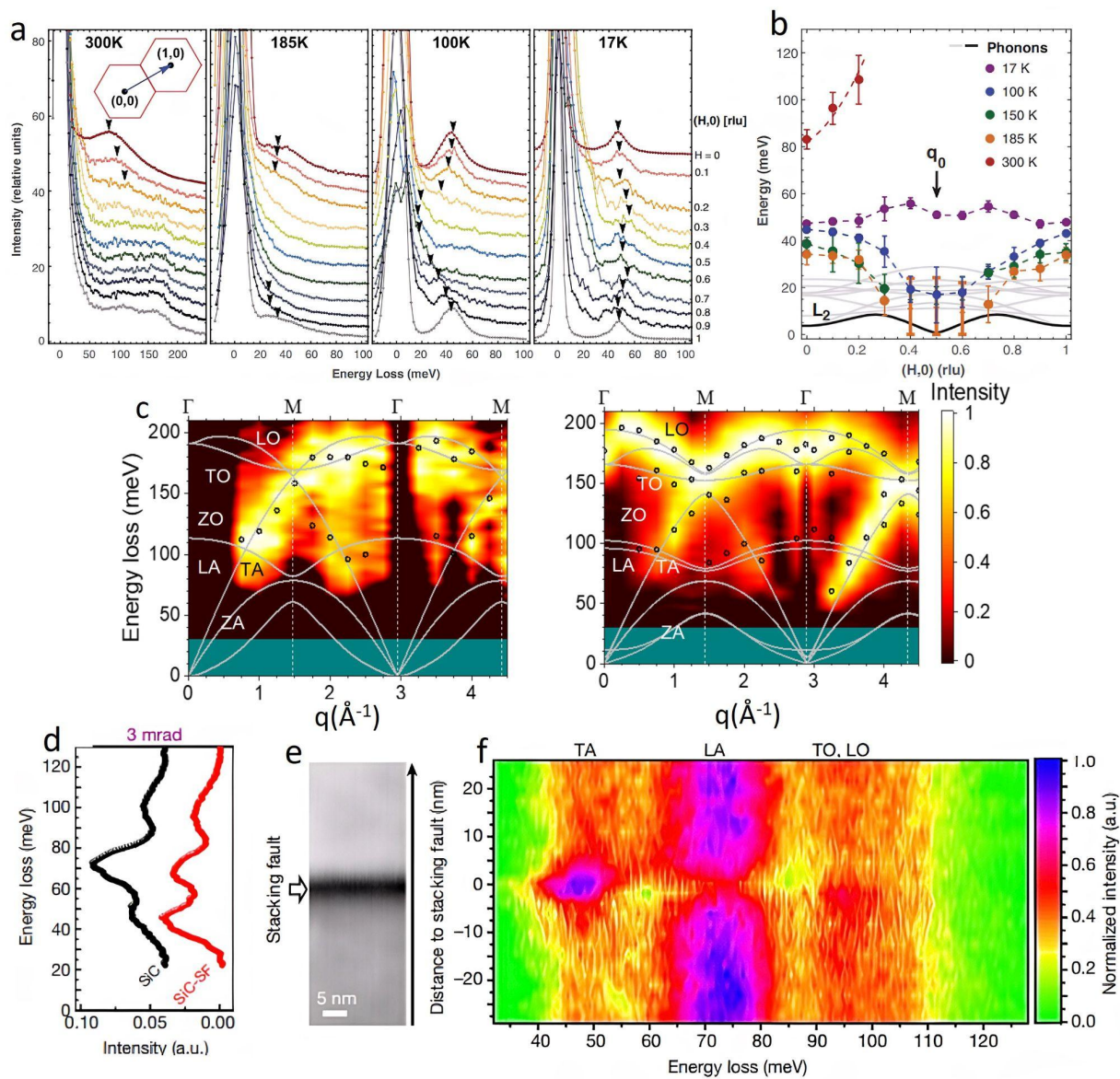

Figure 3. (a) EELS spectra showing momentum dependence of the valence plasmon in TiSe$_2$ for different temperatures. (b) Summary of the momentum dependence of the soft plasmon mode in TiSe$_2$. Dispersion curves along the (1, 0) momentum direction were determined by fitting raw EELS spectra. The plasmon mode behaves like the soft mode of a phase transition, demonstrating the condensation of electron-hole pairs at $T_C$. (a,b) Adapted from ref. [127]. Copyright 2017 American Association for the Advancement of Science. (c) Pseudocolor intensity maps of multilayer hBN (left) and monolayer graphene (right), constructed from the measured EELS spectra, shown with the simulated phonon dispersion curves (solid lines). Peak positions extracted from the measured spectra are indicated by open triangles, showing a good agreement between the measured and calculated dispersion of phonons. Adapted from ref. [139]. Copyright 2019 Springer Nature. (d) Angle-resolved EELS vibrational spectra of defect-free SiC (black) and of SiC with a stacking fault. (e,f) ADF STEM image of the stacking fault and Line

profile of the angle-resolved vibrational spectra at the X point of the first Brillioun zone across the stacking fault in the direction denoted by the black arrow in (e), showing the spatial distribution of individual phonon modes and the red shift of the localized phonon. (d,e,f) Adapted from ref [138]. Copyright 2021 Springer Nature.

### 3.2. Mapping exciton states and single photon emission in low-dimensional quantum materials using combined CL and EELS

Low-loss monochromated EELS (mono-EESL) studies have successfully probed excitons in a series of Rydberg states in multilayers, single layers, lateral heterostructures, and vertically stacked twisted heterostructures of transition metal dichalcogenides (TMDCs) with nanometer precision from cryogenic to room temperature.[140–142] Using hyperspectral EELS imaging, the spatial distribution of excitons has been resolved at five different positions across the interface of a $MoS_2$-$MoSe_2$ lateral heterostructure (Figure 4a). The excitonic signature variations including peaks broadening are further linked to chemical composition variations using low-loss mono-EELS in conjunction with core-loss EELS.[140] Meanwhile, the dispersion relation of excitons has been directly measured by q-EELS for understanding exciton dynamics.[141] The study reveals a parabolic dispersion relation of valley excitons in freestanding monolayer $WSe_2$ along Γ-K direction as shown in Figure 4b. Additionally, subgap exciton states are probed and attributed to defect localized excitons (Figure 4c).

Electron-induced CL (electron in-photon out) spectroscopy has been used to measure optical emissions of excitons and trions (i.e. charged excitons) in single layer TMDCs.[13,143] In contrast to the broad excitonic features shown in EELS spectra in Figure 4a-b, excitonic transitions give rise to well-defined peaks in the CL spectra. Therefore, CL spectroscopy enables accurate determination of the excitonic transition energies. A CL study of $WS_2$ encapsulated in hBN has shown a local modulation of both emission energy and intensity of exciton and trion in immediate proximity (within a few nanometers) of inhomogeneities such as dielectric patches and wrinkles in the $WS_2$ encapsulated in hBN as shown in Figure 4d-f.[13] Furthermore, an enhancement in trion emission is observed and correlated with the regions where the charge accumulation is likely to occur.

Note that it is crucial to encapsulate the atomically thin 2D semiconductors on either both sides or one side for enhanced CL and EELS signals.[14,143] For instance, hBN capping reduces the roughness and dielectric disorder and passivates defects on the surface of the TMDCs, significantly reducing the inhomogeneous broadening of the exciton transition.[144] Encapsulated TMDC monolayers show significantly sharper EELS absorption lines in contrast to TMDCs without encapsulation.[145] In CL spectroscopy, the capping forms a hBN/TMDC/hBN heterostructure that increases the interaction volume with the incident electron beam. The e-beam interacts with the hBN to create electron-hole (e-h) pairs that are funneled into the TMDC layers via the heterostructure, resulting in an enhanced optical emission. Additionally, atomic resolution STEM imaging and subnanometer resolution 4D-STEM analysis of monolayer TMDC has been achieved with hBN encapsulation, enabled by the large atomic number difference (Z contrast) and large lattice constant difference between the TMDC and the hBN.[13]

Another intriguing direction is to apply correlative electron-optical spectroscopy techniques to study exciton localization due to defects, strain, and the moiré potential in atomically thin 2D semiconductors and heterostructures. Localized excitons can act as single-photon emission sources for quantum optics applications. In particular, excitons and trions are localized by the moiré potential in vdW homostructures and heterostructures. The localization can be tuned by the twist and lattice mismatch between constituent monolayers. Moiré engineering of excitons provides a promising avenue to realize a controlled array of single-photon emitters.[22,146–148]

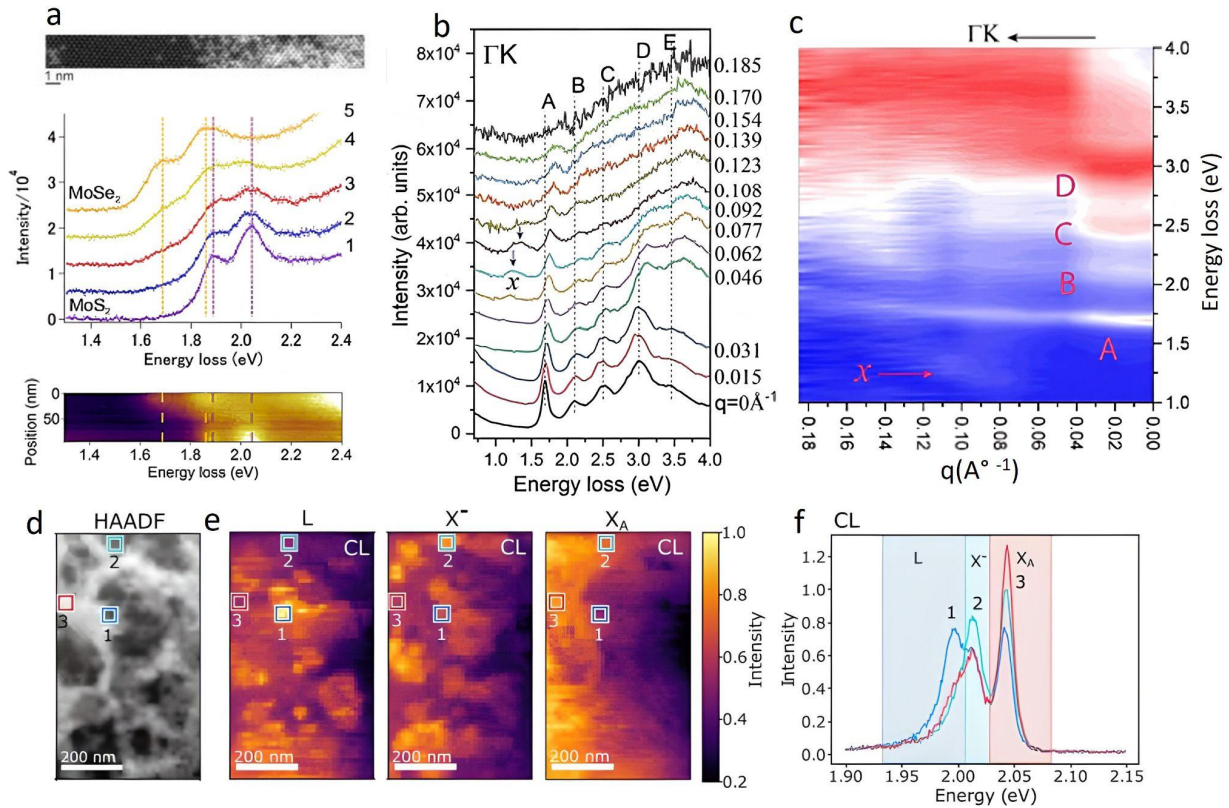

Figure 4. (a) (upper) HAADF image of an interface across a lateral $MoS_2$-$MoSe_2$ heterostructure. (Middle) Five spectra collected at different positions across the interface, showing the variation of excitonic signatures across the interface (lower) 1D EELS map in the spatial direction perpendicular to the interface. (a) Adapted from ref.[140]. Copyright 2015 American Physical Society. (b) Momentum resolved EELS of $WSe_2$ monolayer along Γ-K direction. Peaks marked as x,A,B,C,D,E are attributed to different excitonic transitions. (c) Momentum-Energy diagram showing the dispersion of several exciton states. (b,c) Adapted from ref. [141] Copyright 2020 American Physical Society. (d) HAADF image of an area with local inhomogeneities in a monolayer $WS_2$ encapsulated with hBN layers. (e) TEM CL intensity maps of defect trapped excitons denoted as L (left), trions denoted as X– (center), and free excitons denoted as XA (right), where localized spots are seen for the emission from L and X-. (f) CL spectra corresponding to highlighted regions in (e).(d,e,f) Adapted from ref. [13]. Copyright 2021 American Chemical Society.

## 3.3. Probing single photon emission in color-center-hosting nanostructures using CL-based correlation techniques

Point defects embedded in the crystal lattice of semiconductors and insulators can behave as near ideal emitters with atomic-like states. These quantum defects, also referred to as color centers, can be coherently excited and manipulated, thus acting as spin-qubit and single photon emitters. Color centers such as the nitrogen-vacancy (NV) center in diamond and silicon carbide as well as single photon emitters in hBN attract vast research interest due to their high stability, long coherence times, and narrow spectral linewidths.[149–152]

CL-based techniques, including SEM-CL and S/TEM-CL, provide sub-wavelength resolution with the capability to excite and detect emission in a wide spectral range from ultraviolet (UV) to near-infrared (near-IR). Moreover, CL spectroscopy shows high sensitivity to probe single defects.[153] Using CL, the NV-centers in diamond are spatially and spectrally resolved.[154–156]. Furthermore, CL studies have demonstrated new types of single photon emitters in hBN in the UV and visible regime.[157–159] Additionally, a recent CL study of moiré metamaterial (twisted hBN flakes) has demonstrated a new type of UV single photon emitter that can be tuned and manipulated by the twist angle.[160]

S/TEM-CL spectroscopy in conjunction with high-resolution S/TEM imaging, nano-beam electron diffraction and spectroscopic analysis, enables the correlation of the nano-scale optical properties of the color centers to the local atomic and chemical environment. S/TEM-CL is well suited for probing the effects of local complexities such as crystallographic defects, strain fields, and impurities on the emissive properties of the color centers. In the study by Hayee et al, a correlation of STEM-CL spectral mapping and photoluminescence (PL) spectroscopy of hBN flakes has been used to identify and locate several classes of quantum emitters with different emissive properties as shown in Figure 5a-b. The quantum emission was attributed to four different classes of localized radiative defects present in hBN flakes.[159]

When CL signals are analyzed via photon-correlation spectroscopy e.g. second order intensity correlation measurements ($g^2(\tau)$), both excitation and emission characteristics can be measured. First, the $g^2(\tau)$ autocorrelation function has been used to ascertain single-photon emission from single optical emitters. To calculate $g^2(\tau)$, emitted light in CL is analyzed using a Hanbury, Brown and Twiss (HBT) intensity interferometer coupled with time correlated single-photon counting techniques. The HBT interferometer has been integrated into CL systems for the detection of the CL photons emitted from single photon emitters.[155] Figure 5c, shows a schematic of a setup for measuring the CL $g^2$ function. A SEM image of the molecular epitaxy beam (MBE) grown InGaN/GaN QW nanowires (NW) and corresponding $g^2(\tau)$ maps are shown in Figure 5d-e. The $g^2(\tau)$ map of the electron excited emission however shows a bunching peak with $g^2(0)$ far above 1, which is in stark contrast to the $g^2(\tau)$ measurement using photoexcitation, as shown in Figure 5e ( left panel vs. right panel).[161–163] At higher beam currents, the bunching amplitude is reduced owing to a shift of the overall photon statistics towards the near Poissonan photon-statistics of the electron beam. The plasmon diffuses and decays, resulting in the excitation of multiple color-centers and the simultaneous emission of multiple photons. The shape of the CL $g_2(t)$ function, both for bunched and anti-bunched signals, has been used for quantifying the emission lifetime of the single photon emitters.[162,164]

Ultimately, S/TEM-CL can be combined with atomic resolution S/TEM imaging and 4D-STEM for direct imaging of the atomic structure of the color centers. Figure 5f-h shows the CL spectral mapping has been correlated to 4D-STEM strain mapping around the SiV quantum emitters to identify the strain effect on the spectral variation of the emitters. In addition, atomic-resolution core-loss S/TEM-EELS imaging can be applied to unravel the electronic charge states of the atomic defects. The key findings on atomic configurations and the charge states of the color centers will provide fundamental understanding of various quantum defects and their rich quantum-optical properties, which is essential for defect engineering of the quantum emitters for various quantum applications.[156]

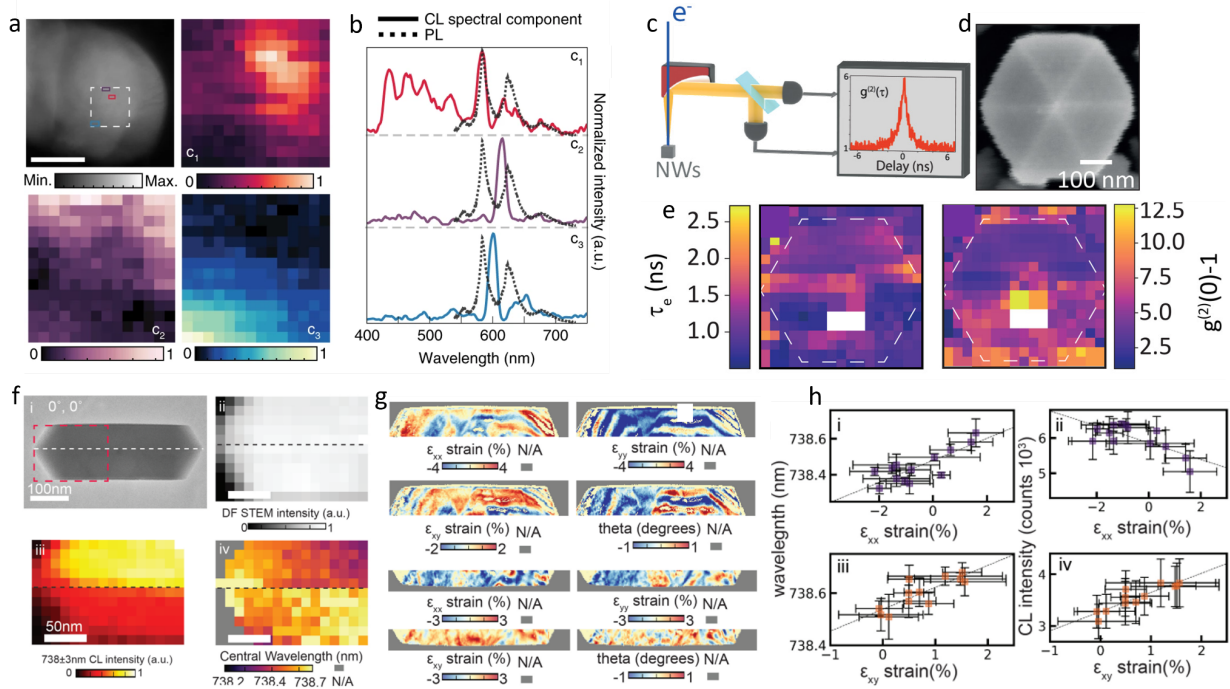

Figure 5. (a) STEM–HAADF image of a region hosting several quantum emitters in a hBN flake and CL spectral weights of the white marked region. Scale bar in the HAADF image is 100 nm and each pixel in the CL spectral map is 5 nm. The red, purple and teal coloured regions of interests (ROIs) indicate the positions of the C1-C3 spectral components. (b) The CL component spectra (solid lines) and normalized PL spectra (black dotted lines) of the C1-C3 regions (marked on STEM-HAADF image (a)). (a,b) Reproduced with permission from ref. [159] Copyright 2020 Nature publishing group. (c) Setup for measuring the CL/$g^2$ function. (d) SEM/CL image of InGaN/GaN quantum wells in GaN nanowires. (e) Maps of lifetime $\tau_e$ (Left) and $g^2(0)-1$ (Right) derived from $g^2(\tau)$ fits. (c-e) Reproduced with permission from ref.[165]. Copyright 2018 American Chemical Society. (f) i TEM image ii: DF stem counts, iii: 738± 3 nm summed intensity, iv: Central wavelength of lorentz fit of SiV- quantum emitters. (g) Strain maps of the top and bottom crystallite divided by a twin boundary i-iv: xx strain, yy strain, xy shear,

rotation obtained from 4D STEM analysis. (h) i-ii: ZPL wavelength vs yy strain and xy shear. iii-iv: ZPL intensity vs yy strain and xy shear, for CL data taken along the purple line that crosses the middle boundary of the particle. Copyright :unpublished work of Jennifer Dionne's group.[156]

## 3.4. Understanding nanoscale chiral light-matter interactions in nanostructures using circular-polarization resolved CL

Circular-polarization resolved CL offers the capability to unravel chiral light-emitter interactions at the nanoscale.[166,167] Chiral light-matter interactions realized in valleytronic materials and photonic structures offering new functionalities and applications for quantum information processing and quantum optics. In valleytronic 2D materials, such as the TMDCs, excitons are coupled to the two energy degenerate (conduction and valence band edges) but inequivalent valleys in the Brillouin zone.[168] The valley degree of freedom can be manipulated via circularly polarized light (CPL) due to the valley-selective optical selection rule.[169] The chiral coupling between quantum emitters and phononic nanostructures can lead to propagation-direction-dependent emission, scattering and absorption of photons, enabling non-reciprocal single photon devices.[170–172]

In photonic nanostructures, the incident electron beam can coherently excite plasmonic or optical Mie modes, leading to a circularly polarized near-field optical response. The CPL-selective excitation, combined with the ability to analyze the CPL emission, enables the circular polarization resolved CL. Probing photonic nanostructures with circular polarization resolved CL holds great promise for the valleytronic study of 2D materials where excitons at different valleys can be selectively excited and detected on deep subwavelength scales. The circular polarization resolved CL technique can provide direct access to critical information in the context of valleytronics including the diffusion and nanoscale dynamics of various valley specific excitonic states such as neutral excitons, trions and bound excitonic states.[170–172]

## 3.5. Measuring nanoscale thermal transport and coupled excitation dynamics in quantum materials

Complementary to EELS and CL, electron energy gain spectroscopy (EEGS) is another electron spectroscopy technique that can be used to explore a range of quantum properties including retardation effects and sub-fs dynamics of relativistic electrons. EEGS, also commonly referred to as Photon Induced Near Field Electron Microscopy (PINEM) combines the unprecedented spectral resolution of a photon probe with an ultrahigh spatial resolution of S/TEM.[102,173,174] This pushes the spectral resolution limit that is otherwise posed by the full width at half maximum (FWHM) of the zero loss peak (ZLP). EEGS is experimentally done using a pump-probe approach in a TEM equipped with optical illumination and based on a pulsed laser set up; however, recently it has been demonstrated that EEGS is feasible under continuous wave (CW) illumination with a high enough intensity.[103,122,175,176] The details on the pump-probe set up and time-resolved spectroscopy is discussed in the dynamic and ultrafast transmission electron microscopy (UTEM) section.

In PINEM, energy gain (left side of ZLP) and energy loss (right side of ZLP) spectra appear symmetrically at both sides of ZLP. The electron energy gain process includes various stimulated photon absorption processes such as electron-phonon coupling and electron-plasmon coupling. EEGS can be combined with energy filtered TEM (EFTEM) imaging offering direct and background free imaging of the evanescent electromagnetic fields such as near-field distribution of surface plasmon resonances (SPRs).[102,122] Furthermore, combining stimulated EELS and EEGS, one can precisely measure the local temperature in the illuminated region based on the ratio of electron energy loss and gain (emission and absorption). Examples of studies that have been done leveraging EEGS capabilities in quantum materials include understanding the retardation effect in MgO nanocubes, phonon-electron coupling dynamics in MgO,[133] direct measurement of temperature dependent heat dissipation in hBN, loss and gain of the optical-phonon modes in hBN flakes,[177] and plasmon-electron coupling in plasmonic nanostructures including Ag nanorods and Ag nanoshells.[103,176,178–182]

## 4. Retrieving phase information and imaging in quantum materials using four-dimensional STEM (4D-STEM) / momentum-resolved STEM

To form an image in STEM mode, at each probe position I(x,y), the transmitted signal is collected, counted, and integrated into a value per pixel position in reciprocal space within a range of scattering angles defined by the STEM detector. In conventional STEM imaging, multiple monolithic and single-pixel detectors with annular geometry (ring or circular) are available such as bright-field (BF), annular dark-field (ADF) and high-angle annular dark-field (HAADF) detectors. While BF detectors capture the signal from unscattered electrons to very low angle scattered electrons (<0-10 mrad), high angle inelastically scattered electrons can be captured using ADF or HAADF for a given range of scattering angles that is 10-50 mrad for ADF detector and angles> 50 mrad for HAADF detector. Among all STEM imaging modalities, HAADF also known as Z-contrast imaging is the most commonly used STEM imaging technique due to its ease of interpretation. The Z-contrast images are based on the incoherent signal from thermal diffuse scattering (TDS).[183,184]

With all the imaging and spectroscopy capabilities that STEM can offer, the intensity variation in the detector plane is neglected in conventional STEM as only a single value per probe position is recorded. The intensity modulation and oscillations of the scattered probe contain valuable phase information. The phase information, if retrieved, can open doors to quantitatively measure inter-atomic strain, internal electric field, and localized charge distribution including ionicity around defects, interfaces, and boundaries. This further lends itself to additional information on individual chemical bonding and simultaneous imaging of heavy and light-Z elements in complex quantum materials. In 4D-STEM, also referred to as momentum-resolved STEM, the focused probe rasters across the sample and for each probe position of $x,y$ in real space, a corresponding diffraction pattern ($k_x,k_y$) in reciprocal space is acquired. Therefore, the final 4D dataset contains all $x,y,k_x,k_y$ information. In 4D-STEM, nearly all electrons are collected after interacting with the specimen. A 'complete' set of phase and amplitude information about the specimen is encoded in 4D-STEM datasets and can be reconstructed via various phase retrieval methods[185] or virtual detectors that are discussed in this section.

Recent advances in detector technology, computational power, and data analysis are key for acquiring data quickly and retrieving the full phase information from the large 4D-STEM datasets. High-performance electron detectors are beneficial in order to record a high quality diffraction pattern at each probe position and to acquire the full distribution of the transmitted probe. A detector with high detective quantum efficiency (DQE), high dynamic range, high signal-to-noise ration (SNR), and high read-out speed (preferably sub-ms per frame) is required for 4D-STEM data acquisition. Regarding high dynamic range, it is desirable to simultaneously record strong scattering objects and weak scattering objects, corresponding to bright and faint features in the convergent beam electron diffraction (CBED) pattern. The two main types of pixelated direct electron detectors that resemble a Cartesian grid are hybrid pixel array detectors (PAD) and monolithic pixel sensors (MAPS).[185–189] Various post-processing and reconstruction algorithmic approaches, including ptychographic phase retrieval algorithms, can be applied to extract the phase information embedded in 4D-STEM datasets, a capability that is lost in conventional STEM.[183,190]

In the following sub-sections, we will discuss how, by mining the 4D-STEM datasets, we are able to accelerate discoveries in the field of quantum materials. In particular, we will demonstrate case studies across five major 4D-STEM capabilities and provide examples over a range of quantum materials including TMDCs, complex oxides, and perovskites. In these material systems, 4D-STEM has been applied to extract information about localized charge distribution, chemical bonding, polarity and internal field, and inter-atomic strain mapping.

### 4.1. Imaging local crystal structure in weakly scattering and beam sensitive quantum nanostructures using dose-efficient imaging

4D-STEM is a powerful technique for elucidating the complex structure of weakly scattering and beam sensitive materials which otherwise are extremely challenging, and in some cases impossible, with conventional STEM imaging modalities.[191] Weakly scattering crystalline nanostructures consisting of low-Z elements such as graphene and carbon nanotubes are challenging materials to image mainly due to minimal image contrast and beam-induced structural evolution. Using 4D-STEM, it is possible to map heterogeneities in such weakly-scattering systems.[191,192] Beam-sensitive materials refer to material systems with inherent sensitivity to electron beam irradiation. In other words, the sample undergoes rapid structural changes and decomposes upon incidence of the electron beam. The unique combination of fast-data acquisition via direct electron detectors and ptychographic phase reconstruction offers significant enhancement in the signal-to-noise ratio and provides unique opportunities for dose-efficient imaging via electron-dose controlling. Therefore, data acquisition is possible with reduced electron exposure. In addition to atomic and molecular structure determination, 4D-STEM datasets can be mined to identify the local crystal orientation, and measure π−π stacking of organic molecules as shown in Figure 6a,[193] domain size, and degree of molecular ordering. It can further be used to identify the nanocrystalline regions in molecular thin-films.[194]

## 4.2. Measuring internal electric field and charge density distribution of quantum heterointerfaces and single defects

4D-STEM also enables direct visualization and quantitative measurement of the interatomic electric and magnetic field (Lorentz configuration). The beam displacement or deflection for each probe position can be extracted from 4D-STEM datasets. In other words, the electron beam deflection appears as a lateral shift compared to the center of mass (COM) vector maps for each CBED pattern probe position in the acquired 4D-STEM dataset. The beam deflection is correlated to the momentum transfer at each STEM probe position. In adequately thin specimens, with the assumption that the probe intensity does not vary with thickness, the momentum transfer or change in the momentum is correlated to the average electric field induced beam-shift for each probe position (inversely proportional).[195–197]

4D-STEM enables real-space charge-density mapping at atomic resolution. Based on Gauss's law, the localized charge density of the electric field can be calculated.[195,196,198] With this, it is possible to measure the induced electric field surrounding single defects and charge accumulation at an insulator-ferroelectric interface, directly observe two-dimensional electron gas (2DEG) and real-space charge-distribution in perovskites, and identify regions of steady-state negative capacitance in complex oxide superlattices,[199,200] as well as direct imaging of electric field in proximity of quantum defects.[201] Figure 6b shows the Sub-Å map of charge distribution and polarization induced 2DEG at a ferroelectric/insulator heterointerface ($BiFeO_3$/$TbScO_3$).[199]

## 4.3. Simultaneous imaging of multicomponent nanostructures composed of light and heavy elements

4D-STEM has additional functionality in allowing simultaneous Z-contrast imaging in multicomponent nanostructures, thereby enabling simultaneous imaging of light and heavy elements. Figure 6c, shows a comparison of a false colored HAADF-STEM image and a ptychographic reconstructed phase image of a 2D-$MoS_2$ monolayer via 4D-STEM. The significant scattering cross-section variance between high and low-Z elements in complex nanostructures, such as in 2D-$MoS_2$ monolayers with light-Z dopants, is the limiting factor in conventional Z-contrast STEM imaging. The electron beam induced nanopores are partially covered with carbon. The carbon lattice is resolved in the phase image, and, therefore, the regions covered with an ultra-thin layer of carbon are identified in the presence of Mo and S elements. This technique can provide direct insight about individual light element dopants on sublattice sites (interstitial, substitutional) as well as heterovertical interfaces, surface adatoms, or surface adsorbate.[202,203]

## 4.4. Leveraging ptychography reconstruction and lattice-distortion mapping in low-dimensional materials using nano-beam diffraction

Another major 4D-STEM application is in quantitatively mapping local lattice distortions with high precision down to the picometer (pm) length scale. Strain and lattice distortions are derived via ptychography reconstruction and by measuring the strain induced shifts in the position and in the center of mass (COM) of discs in nano-beam electron diffraction (NBED) patterns in ultrathin quantum materials such as atomically thin 2D materials.[184,204,205] It is worth noting that non-overlapping diffraction (Bragg) discs are referred to as NBED patterns.

4D-STEM has been used for a range of quantum materials, from graphene[192,206] to halide perovskites[207] and nanocatalysts.[208] to identify the localized and delocalized strain. In a core-shell nanostructure, the lattice distortion across a heterointerface can strongly mediate the functional properties such as band gap or catalytic activity. In 2D materials, 4D-STEM is used to identify the in-plane and out-of-plane structural distortions, such as the variation in interlayer spacing in multi-layer 2D-materials[206,209] or measuring the twist direction and angle in graphene[192] to understand the superconductivity transitions in such systems. In halide-perovskites, using 4D-STEM, one can directly measure the octahedral rotation angles.[207] Figure 6d presents an example of quantitative strain maps in a single nanoparticle of rhodium@platinum (core@shell) core-shell structure with sub-pm precision. The strain maps are derived from 4D-STEM NBED patterns revealing the strain variations as a function of the distance from the Rd/Pt heterointerface.[208]

## 4.5. Virtual STEM imaging

Figure 6e shows a 4D-STEM dataset that has been acquired from a complex oxide sample $SrTiO_3$ along a [001] direction.[189] A full set of STEM images analogous to various conventional STEM images including (BF, ABF and ADF) has been retrieved through selective masking of the position averaged convergent beam electron diffraction (PACBED) pattern acquired via 4D-STEM. The size, shape, angle, and position of the virtual mask, which is also referred to as "virtual detector", or "universal detector", is quite flexible. Compared to conventional STEM imaging, using 4D-STEM virtual detectors can simultaneously derive all equivalent conventional STEM (BF, ABF, ADF etc.). In addition, any other arbitrary shape of virtual apertures can also be applied to mask an area of interest on PACBED and generate images beyond conventional STEM imaging.[186]

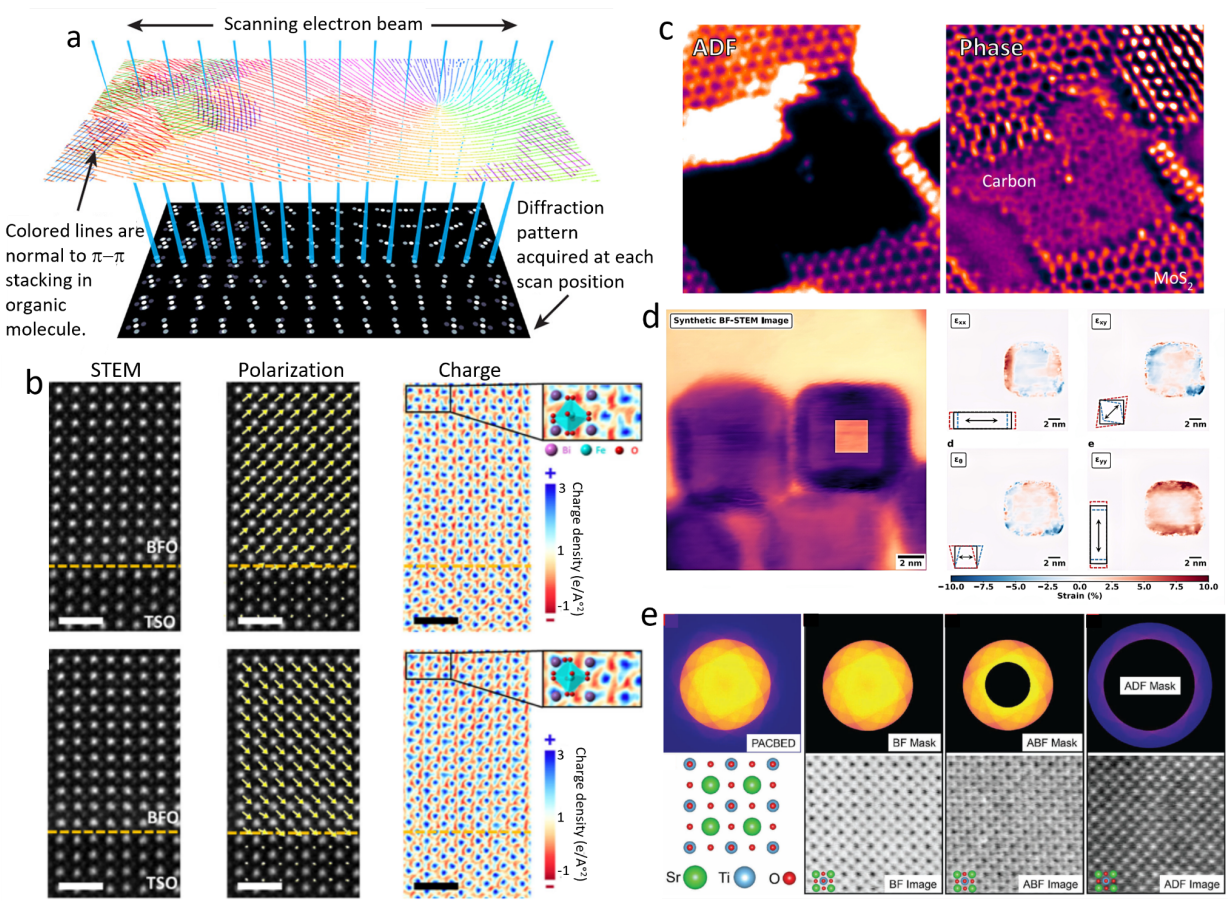

Figure 6. a) 4D-STEM datasets can be analyzed to identify the local crystal orientation. 4D-STEM datasets are mined to measure π−π stacking of organic molecules (polymeric chains), domain size and degree of molecular ordering. Adapted from ref.[191]. Copyright 2021 American Chemical Society publishing group. b) Measuring the internal field and local charge distribution. Spatially resolved maps of local-charge distribution, internal field and local polarization at the BiFeO3/TbScO3 ferroelectric/insulator heterointerface. Adapted from ref.[199]. Copyright 2021 Nature Partner Journals publishing group. c) Simultaneous imaging of light and heavy Z elements. A comparison of a false color HAADF-STEM image and a ptychographic reconstructed phase image via 4D-STEM for 2D-MoS$_2$ monolayer. The carbon lattice in the carbon covered regions are resolved in the phase image and an ultra-thin layer of carbon is identified in the presence of Mo and S elements. Adapted from ref. [203]. Copyright 2019 American Chemical Society publishing group. d) Inter-atomic strain mapping. Quantitative strain mapping in a rhodium@platinum (core@shell) nanocube with sub-picometer precision. The strain maps are derived from 4D-STEM NBED patterns and single unit cell sizes as a function of the distance from the Rd/Pt heterointerface. Adapted from ref.[208]. Copyright 2020 American Chemical Society publishing group. e) Virtual imaging using position averaged convergent beam electron diffraction (PACBED). A full set of STEM images analogous to various conventional STEM images including (BF, ABF and ADF) in SrTiO$_3$ along a [001] direction has been retrieved through selective masking of the position averaged convergent beam electron diffraction (PACBED). Adapted from ref.[189]. Copyright 2020 Wiley publishing group.

# 5. Exploring reversible and metastable dynamics of quantum excitations using ultrafast transmission electron microscopy (UTEM)

The evolution of time-resolved TEM is directly linked to instrument advances over the last decade. One major breakthrough in the development of time-resolved TEM came with the use of ultrashort pulsed laser sources enabling sub-nanometer/femtosecond spatio-temporal resolution.[210–214] In addition, time-resolved TEM has benefited significantly from the same instrumental advances discussed in the previous sections (aberration correctors, direct electron detectors). Understanding the dynamics of involved elementary processes down to the atomic and molecular level had been impossible before the early 2000s and the pioneering work of O. Bastanjoglo at TU Berlin.[215]

Time-resolved TEM can be considered as a direct extension of conventional TEM with expanded capabilities into the time domain, allowing for analysis of ultrafast structural dynamics and transformations with sub-nm spatial resolution and ps-fs timescale.[210–214,216–220] Time-resolved TEM enables the identification of intermediate states that occur during complex transitions thereby providing the basis for a solid understanding of fast dynamics e.g. the displacement of atoms and formation of individual structural point defects through irreversible atomic-bond breakages.[221] "Exactly like someone would ask a magician to slow down his movements and break them into a sequence of elementary gestures to understand a card trick, electron microscopists have long dreamed of being able to excite samples and acquire images, diffraction patterns, or energy spectra after a controllable delay to elucidate their dynamics."--Arnaud Arbouet, Giuseppe M. Caruso, Florent Houdellier.[210] Time-resolved TEM also opens new avenues for investigating structural and chemical properties of beam sensitive specimens including soft materials or quantum nanostructures based on low Z elements by avoiding irreversible electron beam damage. The pulsed imaging mode enables researchers to use significantly lower electron dose via temporally controlled electron dosage.[222,223]

Time-resolved TEM has additional applications that originate from electron/photon interactions. The interactions between ultrashort electron pulses and ultrashort laser pulses opens a door for new discoveries in the nanophotonics field by exploring the linear and non-linear optical response of nanophotonic and optical nanostructures.[224] Explorations range from electron-generated carrier dynamics, such as excitation lifetime, to identifying the decay pathways of excited states with sub-nm spatial resolution and fs-ps temporal resolution.[225–228]

To enable time-resolved spectroscopy and imaging in electron microscopes, various pump-probe configurations have been implemented in EM platforms. During a pump-probe experiment, the sample is probed at a fixed delay after a pump and the change in the signal is monitored. The feasibility and integration of pump-probe experiments inside a TEM have been evaluated [229] during early pioneering works at the Technical University Berlin,[214] Caltech [220] and Lawrence Livermore National Lab.[230]

The electron probe pulses are synchronized with laser-driven events induced in the sample. Individual electron pulses are generated via laser-triggered photoemission from a photocathode. Time-resolved electron microscopy and spectroscopy based on a pump-probe approach can be operated in two modes: either single-shot or stroboscopic. The synchronized electron pulses are accelerated in the electron microscope column and probe the sample. Additionally, the sample is illuminated with a laser pulse referred to as the sample pump beam, with a controlled delay time relative to the electron probe pulse.[211–213,231,232]

In both single-shot and stroboscopic mode, the TEM column is modified to allow the time-resolved laser access to both electron gun and specimen. However, the employed laser system in single-shot mode and stroboscopic modes are significantly different. The single-shot mode is commonly referred to as high-speed TEM or dynamic TEM (DTEM), whereas stroboscopic mode is referred to as ultrafast TEM (UTEM).[233] Both are considered to be time-resolved TEM, and it is important to note that the following sections will refer to all functions as UTEM or ultrafast electron microscopy (UEM).

Using a fs laser as an excitation source, the stroboscopic approach attains spatial resolution down to the atomic level and sub-picosecond temporal resolution by suppressing the influence of Coulomb repulsion. In stroboscopic mode, millions of low-intensity and ultra-short fs electron pulses (electrons/pulse: 1-10) from identical pump-probe events are integrated to make the final image (pulses/image>$10^5$). In single-shot mode, upward of a million electrons are released in a single intense multi-electron bunch/pulse (electrons/pulse~$10^5$-$10^7$). The temporal resolution is limited to the nanosecond regime due to space-charge effects and a few tens of nanometers in spatial resolution (~10-20 nm).[220,234]

Overall, the stroboscopic mode is well-suited for studying reversible physical phenomena. Since electronic excitations and vibrational transitions are reversible at a fixed frequency and phase, the stroboscopic mode can contribute significantly to the field of nanophotonics and quantum science. The single-shot approach can be used to capture transient states in irreversible laser-induced microstructural processes such as sublimation, plastic deformations, and lattice expansions in addition to temporally resolved phase transformation and nucleation/growth of new phases in a matrix. The major disadvantage of single-shot mode is the limited spatial resolution and the major downside to the stroboscopic is its restriction to repeatable processes.[210,220,234]

**5.1. Probing spatially dependent collective excitations dynamics in quantum nanostructures using UEM**

Figure 7a shows the direct probing of strain-wave dynamics in a freestanding ultrathin 2H-$MoS_2$.[225] This study highlights the strong impact that local lattice distortions, in this case step edges with edge-atom relaxations, can have on coherent phonon dynamics. Strain waves arise from atomic-scale structural distortions and induced force fields. Spatially dependent phonon dynamics as a function of the distance to the marked step-edge are quantified for two regions of interest (ROI1 and ROI2) marked on the TEM micrograph. A significant softening of the phonons is observed (yellow and red graphs) in the vicinity of the step edge followed by a plateau in the phonon frequency extending for several hundred of nanometers.[235] In similar studies, thickness-dependent phonon dynamics in layered quantum materials of 1T-$TaS_2$ and 2H-$MoS_2$ are probed in real space.[236] Similarly, ultrafast reciprocal space imaging (ultrafast convergent electron beam diffraction) is employed to study the strain dynamics in a single-crystalline graphite membrane.[237]

Adding the spectral domain to time-resolved S/TEM can contribute significantly to the in-depth understanding of emerging quantum materials.[238–240] A combination of EELS and UTEM was used to measure Bloch modes in dielectric photonic cavities and investigate the coherent electron-cavity-photon interaction. The ultrafast dynamics of the high-quality (Q) photonic cavities are measured for a wide range of wavelengths, incidence angles, and for both polarization modes: transverse magnetic (TM) and transverse electric (TE).[241] Figure 7b shows the reconstructed band structure with sub-10nm spectral resolution in a 20nm $Si_3N_4$ photonic crystal containing arrays of high Q cavities (Figure 7b). The photonic band structure is measured for a range of pump laser wavelengths modulated between 525-950 nm and incident angles ranging between 0-24.4°. The corresponding Bloch modes, the 2D projection of the electric field spatial distribution, is directly visualized for data points marked on the top panel for both TE and TM (lower panels).[241] Figure 7c and 7d highlight how ultrafast electron microscopy(EM) in conjunction with EELS can be employed to directly resolve the temporal evolution of plasmon resonances and evanescent light field in plasmonic nanostructures. Figure 7c shows the plasmonic field enhancement in the vicinity of an Ag nanowire on a $Si_3N_4$ substrate before and after photoexcitation. The direct imaging of the plasmonic maps is enabled by subtracting of the EELS spectra before and after photoexcitation.[238] In Figure 7d, the polarization-dependent plasmonic response of an Ag nanowire can be resolved via ultrafast EM with fs temporal resolution on the nm length scale.[242,243]

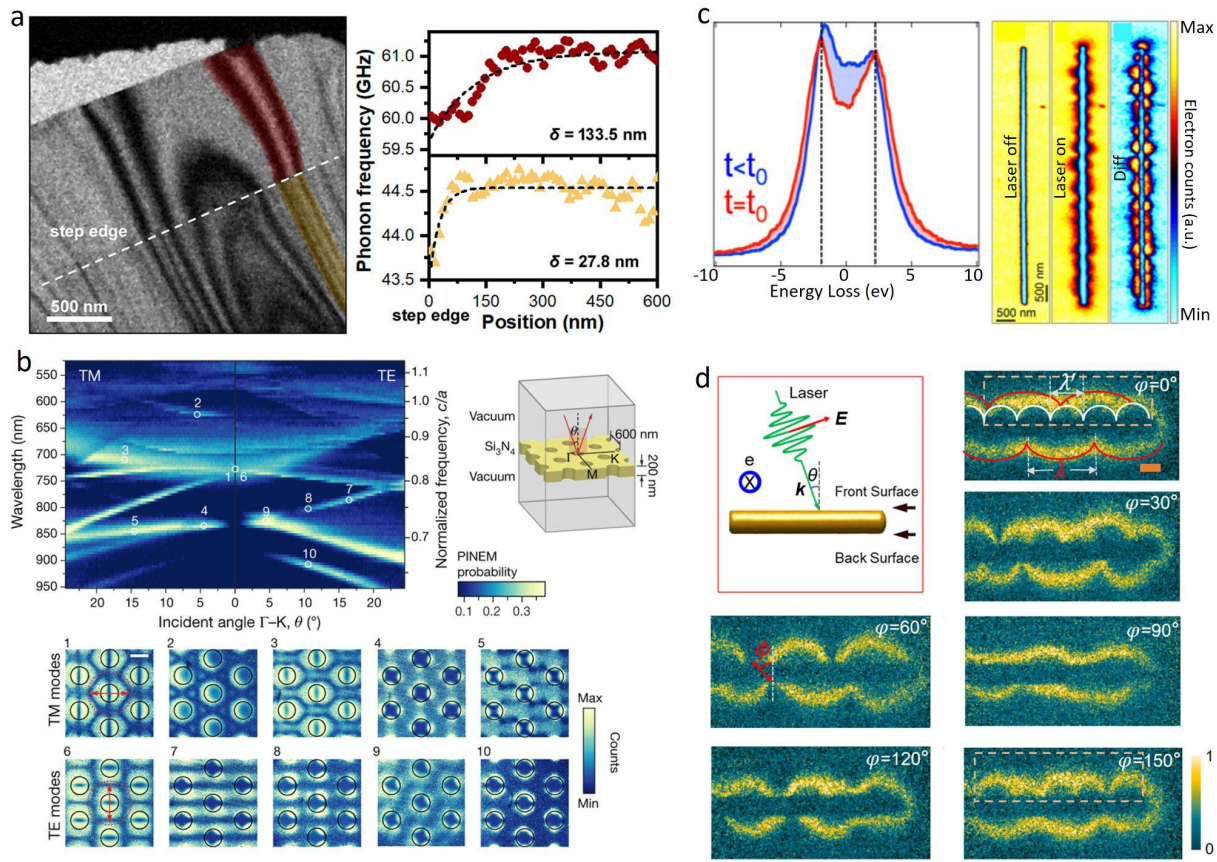

Figure 7. a) Direct visualization of the Bloch modes around photonic cavities in a photonic crystal. Top left: measured band structure for both transverse magnetic (TM) and transverse electric (TE) polarizations by scanning electron beam for incident laser angle of 0°–24.4° (laser line width: 5-10 nm in VIS range) and wavelength range of 525-950nm with 5nm wavelength step size. Top right:Photonic crystal containing photonic cavities and incident pumped laser configuration. Bottom Rows: Corresponding Bloch mode maps of marked data points in the band structure graph for TE and TM modes. Adapted from ref. [241]. Copyright 2020 Nature publishing group. b) Anisotropic Strain-Wave Dynamics and nanoscale phonon softening in 2D $MoS_2$. Left: TEM micrograph of free-standing 2D-$MoS_2$. Right: Variation in phonon-frequency as a function of position (position dependent phonon frequency) for the two (yellow and red ROIs) marked on the TEM micrograph. Positions are measured relative to the step-edge (position 0nm) of the sample. Adapted from ref.[225]. Copyright 2021 American Chemical Society publishing group. c) Imaging the photo-induced plasmonic field in silver nanowire (NW) via swift electrons. Left Spectrum: Acquired EELS spectrum from the silver nanowire before (blue) and during (red) the light excitation at time $t_0$. Right maps: from left to right displays corresponding plasmon maps obtained from the NW with (red), without (blue) and subtraction of red and blue spectrum indicating the photoinduced plasmon map for Ag NW deposited on a $Si_3N_4$ membrane (130 nm diameter, 7.8 μm length). Adapted from ref.[244]. Copyright 2017 American Chemical Society publishing group. d) Polarization dependent nanoscale photo-induced plasmonic field for Ag NW. Top left: Pump-probe configuration setup relative to the Ag nanowire. Corresponding photo-induced plasmonic maps for varying optical polarization(φ), φ ranging between 0°-150°.

Nanowire diameter 110 nm; scale bar 100 nm. Adapted from ref.[242]. Copyright 2021 American Chemical Society publishing group.

## 5.2. Resolving emission dynamics of quantum excitations using ultrafast CL (time-resolved CL)

Ultrafast time-resolved CL has been employed in both scanning electron microscopes and transmission electron microscopes to study the emission dynamics in several different quantum nanostructures such as quantum emitters with nitrogen vacancy (NV) centers,[239] InGaN/GaN core-shell nanowires, GaAs pyramid nanostructures [245] and InGaN quantum wells.[246] Emission dynamics including carrier lifetime, carrier diffusion length, and effective excitation volume are quantitatively measured using time-resolved CL.[247,248]

In quantum emitters, the lifetime of defects and carrier diffusion lengths from NV centers in nanodiamonds are resolved with sub-nanosecond temporal resolution,[249] providing an in-depth understanding of the radiative pathways including excitation and subsequent relaxation of the excited states. Furthermore, offering ps-fs temporal resolution, nm spatial resolution and meV spectral resolution, time-resolved CL in S/TEM is a powerful tool to measure the carrier lifetime modulations as a function of localized microstructural variations, such as local strain due the presence of defects. As an example, exciton dynamics are spatially resolved around a single dislocation in an epitaxially grown GaN on a free-standing GaN substrate.[248] Using time-resolved CL, exciton lifetime, CL intensity, and energy shifts are spatiotemporally resolved as the probe rasters across a dislocation.[248]

## 5.2. Probing nanoscale structural dynamics and light-induced phase transformation with picosecond (ps) temporal resolution

Picosecond time-resolved structural dynamics and phase transitions are highlighted in Figure 8. In the shape memory alloy of $Mn_{50}Ni_{40}Sn_{10}In$, a light-induced martensitic phase transformation and the structural oscillations between a martensitic (MT) phase and an austenitic (AUS) phase are directly visualized.[250] Upon photoexcitation at 100K, a picosecond phase transition from an orthorhombic MT phase to a cubic AUS is confirmed based on the electron diffraction patterns along the [001] zone axis (Figure 8a). The splitting of the diffraction spots (black arrows) are a signature of the MT structure as it differs from the AUS phase. The intensity modulation of the diffraction spots is spatially plotted over the scan span of ~70 pixels (lower left panel) confirming that spot splitting disappears around 13ps (Figure 8b). Furthermore, a photoinduced periodic oscillation between the MT and AUS phases (annihilation and creation of MT domains) originating from the acoustic phonon mode and ultrafast lattice motions is directly observed (Figure 8c and 8d). In Figure 8c, a series of time-resolved images indicate an annihilation of MT domains at 23ps, 66 ps and 110 ps and subsequent reverse transition and creation of MT domains at 46ps and 90ps.[250]

Figure 8e highlights ultrafast dynamics of charge density wave (CDW) phases and light-induced domain wall movements in a free-standing thin film of a 1T-TaS$_2$ TMDC (1T phase Tantalum Disulfide). Employing ultrafast dark-field (DF) electron microscopy, superlattice reflections of

1T-TaS$_2$ in the electron diffraction pattern (EDP) are selectively transmitted via a customized DF aperture.[251] A real space image of the pronounced CDW domains is recorded due to the contrast enhancement of dark field micrographs. A series of spatiotemporally resolved DF micrographs in Figure 8e show the CDW dynamics and transition from a nearly commensurate (NC) phase into an incommensurate (IC) with pronounced dark contrast and back to NC phase with bright contrast. IC domains with dark contrast start nucleating within a few picoseconds from the beginning of the process and become the dominant phase after 0.5 ns while the NC domains start re-forming within a few nanoseconds.[251]

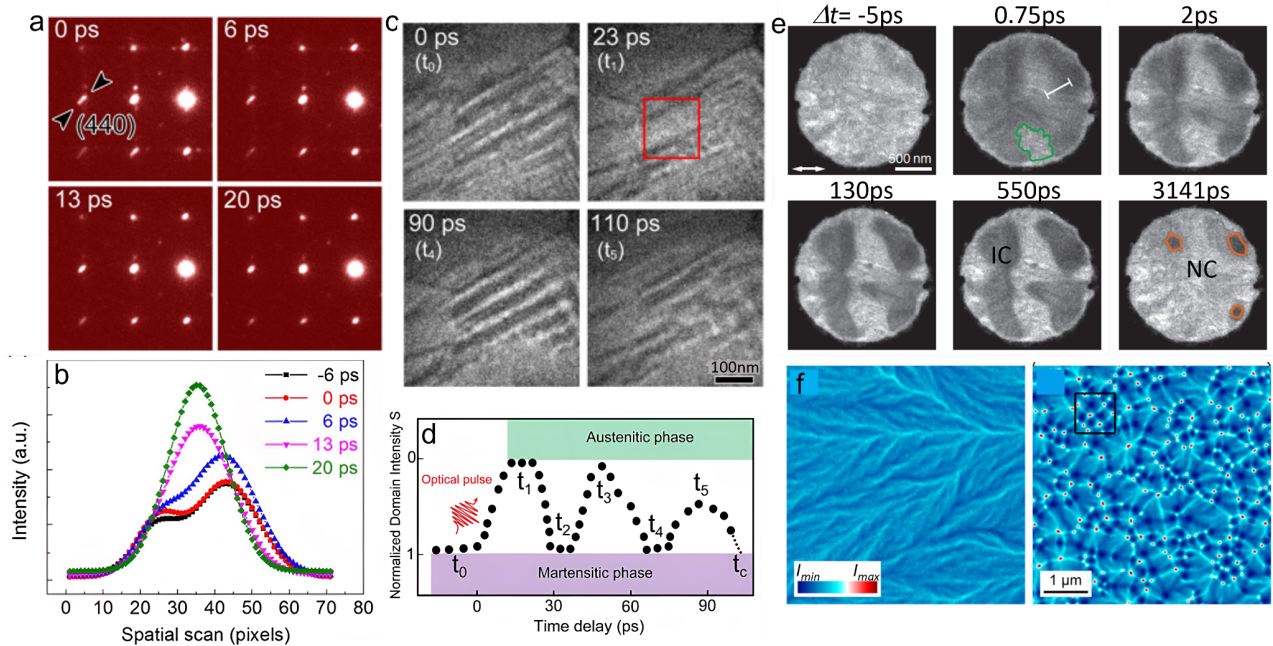

Figure 8. (a,b,c,d) Picosecond time-resolved structural oscillations and photoinduced phase transitions from orthorhombic martensitic (MT) phase to austenite (AUS) phase in Mn$_{50}$Ni$_{40}$Sn$_{10}$In. a) Time-dependent SAED patterns acquired along [001] zone axis at time delays of t= 0, 6, 13, 20 ps demonstrating spot splitting (black arrows). b) The ultrafast changes of spot splitting is spatially plotted for the intensity modulation of the diffraction spots. The spot splitting (signature of MT phase) disappears for t> 13ps. c,d) Series of time-resolved TEM images and schematic of the periodic structural oscillation between MT and AUS phase including a periodic annihilation of MT domains at time delays of t= 23, 66, 110 ps and creation of MT domains at t= 46, 90ps. Adapted from ref.[250]. Copyright 2017 American Physical Society publishing group. e) Series of DF-TEM images displaying ultrafast dynamics of CDW phases and domain wall movements in a thin film of a 1T-TaS$_2$ TMDC. Adapted from ref.[252]. Copyright 2021 Science publishing group. f) Time-resolved Lorentz TEM images of polycrystalline Fe film on Si$_3$N$_4$ substrate demonstrating frozen" metastable magnetic textures using an intense optical pulse the formation of light-induced vortex-antivortex networks. Adapted from ref.[253]. Copyright 2017 American Physical Society publishing group.

## 5.3. Characterizing vortices and topological quasiparticles with time-resolved Lorentz microscopy

Time-resolved UTEM, in combination with Lorentz electron microscopy, is employed to spatiotemporally probe the evolution of nanoscopic magnetic configurations.[254] Recently, ultrafast Lorentz microscopy is used as a novel tool in several studies to address a plethora of open questions related to ultrafast formation and dynamics of magnetic textures.[254,255] Nanoscale magnetic textures, such as vortices and skyrmions are useful for manipulating various couplings and energy transport mechanisms and ultimately altering the functional properties in magnetic nanostructures.[256]

Both stroboscopic and single-shot mode time-resolved TEM have been employed to explore ultrafast light-induced magnetism. Using time-resolved single-shot Lorentz microscopy in conjunction with external perturbations, several novel metastable magnetic textures and far-from equilibrium spin states have been created and characterized.[253,257,258] In Figure 8f, using time-resolved Lorentz TEM, the formation of light-induced vortex-antivortex networks (transition from a ripple texture) in a polycrystalline iron film is demonstrated. The iron film is supported by a silicon nitride substrate in which an intense single shot optical pulse and subsequent local heat pulse is the driving force behind formation of the "frozen" metastable magnetic textures and long-lived changes.[253] In a similar study, with single shot illumination, light-induced vortex switching and ultrafast demagnetization dynamics have been captured in a $Ni_{80}Fe_{20}$ permalloy disc as a high permeability magnetic alloy.[257]

In stroboscopic illumination, femtosecond laser induced heat pulses and subsequent fast quenching are used for writing and recovering the metastable skyrmion phases in FeGe chiral magnets. A series of laser-induced phase transitions from a mixed helical/conical magnetic phase to a metastable skyrmion phase and the magnetization reversal dynamics are captured with fs temporal resolution.[258] In addition to reversible phase transitions in magnetic structures, in FePt films, the photoinduced demagnetization dynamics, including domain wall movements and the degree of coherence at the domain walls, have been directly captured and quantitatively measured with picosecond temporal resolution.[255,259,260] In a similar study, the reversible magnetization dynamics, including nucleation, oscillation, and annihilation of magnetic domain walls, is directly visualized in ferromagnetic Ni thin films.[261] Direct observation of the transient states and understanding the nanoscale magnetization dynamics on ultrafast timescales are crucial for device developments in vortex-based data-storage, spintronics, and quantum computing.[254,259,260]

## 5.4. Enhancing temporal and spatial resolution for sub-cycle quantum physics using highly coherent electron pulses

Radio frequency TEM (RF-TEM) achieves ultrafast temporal resolution through an RF-driven modulation of the electron source and chopping the continuous electron beam into MHz-GHz pulses. With this approach, RF-UTEM overcomes the common challenges of laser-driven UTEM e.g. the constraints on the spatial resolution imposed by the electron pulse quality, coherence, limited repetition rates (kHz range), and beam instability. RF-UTEM, known colloquially as laser-free UTEMs, is designed based on a low-power RF resonant cavity without any major instrument alterations. Using a RF-compressed electron beam, RF-UTEM can generate high quality pulses with increased spatial coherence and brightness as well as tunable frequencies with high repetition rate in the MHz-GHz regime.[211,262–266] By pushing the spatial resolution of UTEM into the atomic scale, it is expected that RF-UTEM will demonstrate significant capabilities in unfolding basic quantum phenomena such as enabling direct observations of atomic motion dynamics and individual atom displacements with fs temporal resolution.

Overcoming the fs temporal resolution limit and pushing the temporal resolution into the attosecond regime is another promising direction that the field of UTEM is moving toward. Attosecond UTEM will open doors to unravel and visualize quantum phenomena with sub-optical cycle durations and directly map the quantum states of electrons. Achieving temporal resolution down to the attosecond regime is recently reported based on a few different configurations.[267,268] It has been shown that a continuous-wave laser induces attosecond short pulses in the TEM continuous electron beam.[268,269] Additionally, by taking advantage of ponderomotive interactions and using a multi-stage compression configuration, attosecond compression of electron pulses was successfully reported. This was achieved by generating picosecond electron pulses and further modulation of the picosecond electron pulses via a fs laser to create attosecond electron pulses. [270–272]

In summary, UTEM has enabled a revolutionary step in quantum science by directly imaging Bloch modes in photonic nanostructures,[241] spin crossover (SCO), charge-density waves (CDW)[252] and polariton wave packets and acoustic phonon dynamics,[225] and by probing light-induced mechanical switching, chirality and angular momentum using electron microscopes.[273] Moving forward, UTEM is a powerful research tool with ultrahigh spatiotemporal resolution combined with meV energy resolution and holds great practical significance for understanding and direct visualization of a wide range of quantum phenomena.[213,274]

## 6. Elucidating ultrafast dynamics of quantum orders using complementary UED, XFEL, and optical spectroscopies

In addition to UEM, complementary spectroscopies, including ultrafast electron diffraction (UED), x-ray free-electron laser (XFEL) based-techniques, and optical spectroscopies, can be used to further elucidate ultrafast dynamics. UED also permits the direct probing of the reciprocal space dynamics in various nanosystems. Such techniques benefit from, but do not require, as high-brightness electron sources as UEM. Further, many UED designs require less sophisticated instruments, such as aberration correctors, electron detectors, and high-accelerating electron guns, while still providing $Å^{-1}$-level momentum resolution, sub-ps temporal resolution, and versatile operation modes.

Femtosecond x-ray scattering and spectroscopy represent another route besides UEM/UED that can access the dynamics of structural, electronic, magnetic degrees of freedom and their coupling. These degrees of freedom are the key to understanding the underlying non-equilibrium physics behind exotic phenomena in quantum materials. Coherent ultrafast x-ray pulses from XFEL facilities provide insights beyond UEM/UED with unprecedented resolution, particularly in time and momentum. It is important to note that this section is intended to show the complementing capabilities of ultrafast x-rays to the electron microscope rather than fully covering ultrafast x-ray techniques and advances, which are thoroughly covered elsewhere.[275–280]

## 6.1. Disentangling quantum orders in rare earth tritellurides and moiré heterostructures using femtosecond UED probes

Besides UEM, the UED technique also permits the direct probing of the atomic-scale dynamics in various nanosystems since its development over two decades ago.[217,281–284] UED does not require as high-brightness electron sources as UEM, but still benefits from the increased brightness. It also has less sophisticated instrumental requirements, while still providing atomic-level views with sub-picosecond temporal resolution. For example, aberration correctors and direct electron detectors are optional with electrons accelerated to energies from sub-keV up to a few MeV. Besides, the relaxed requirement for beam quality allows exploration of versatile operation modes like time-resolved diffraction tomography with a series of sample tilts[285] that has served as the pioneering structural-sensitive probe system for visualizing lattice dynamics upon external perturbations and has found a myriad of applications in material science, chemistry, and biology, among others. Especially in research related to quantum science, the atomic-level insight provided by keV-UED presents a complementary approach to optical probes which are often dominated or complicated by electronic responses.

The recent development of UED facilities with MeV electron kinetic energy further mitigates many challenges in keV UED for quantum science research. The high energies significantly reduce the probability of multiple scatterings, allowing the application of kinematic scattering theory and specific higher-order analyses of diffuse scattering dynamics.[286–288] The Coulomb repulsion between electrons is also reduced by relativistic effects when accelerating the electron to MeV, thus improving the temporal and momentum resolution. For example, the MeV-UED facility at SLAC National Accelerator Laboratory has achieved < 200 fs electron pulses, ~ 0.17 Å$^{-1}$ momentum resolution, and is still undergoing rapid developments.[289] Therefore, these developments have fueled the study of dynamics in many classes of quantum materials. This includes the ultrafast rearrangement dynamics of charge,[290–293] orbital,[293–295], lattice[296], and topological orders [291,297,298] in strongly correlated materials such as complex oxides and TMDCs.

One representative example of quantum phenomena under intensive study is the charge density wave (CDW) order in materials such as the rare-earth tritellurides (RTe$_3$). Due to their rich and well-characterized CDW phase diagram, rare-earth tritellurides serve as a unique platform to study phase competition and non-adiabatic transitions. Such underlying physics remains elusive for many problems in condensed matter research like unconventional superconductivity.[299] Zong *et al*. performed UED measurements on LaTe$_3$, a member of RTe$_3$ with unidirectional CDW ordering to discern its structural modulation upon photoexcitation.[291] By monitoring both Bragg and superlattice peak intensity and width dynamics, the CDW amplitude dynamics were separated from the phase coherence dynamics and found to recover on faster timescales. This different recovery rate was confirmed in two other time-resolved experiments and explained by the presence of topological defects.

UED was also used to reveal the competition of two CDW orders in LaTe$_3$ and their manipulation with light.[290] As shown in Figure 9a-c, photoexcitation weakens the CDW ordering along the *c*-axis, and a different competing CDW order along the *a*-axis emerges. The relaxation dynamics of new CDW peaks and reestablishment dynamics of original CDW peaks are nearly identical at all the fluences, which indicates its origin from topological defect generation upon photoexcitation (Figure 9d). Similar light-induced hidden CDW states are reported with UED in another RTe$_3$ compound CeTe$_3$.[300] UED also facilitates detailed studies of transition dynamics in LaTe$_3$, such as critical slowdown and increasing relaxation time of certain observables, e.g, CDW suppression, near phase transition boundaries [301] and explorations of CDW materials with exotic phase transition behaviors.[302]

In the future, ongoing efforts leading to improved UED momentum resolution[303] and temporal resolution[304] could open tantalizing opportunities in the studies of novel phases of quantum matter in other systems such as moiré heterostructures[28,29] and magnetic orders with long periodicity.[305–307]

## 6.2. Probing quantum phenomena in correlated electron materials using coherent ultrafast x-ray pulses with enhanced momentum and temporal resolution

The development of ultrafast x-ray probes dates back over three decades, enabled by developments in high-intensity amplified optical pulses[308–310] and using x-ray fluorescence from plasmas[311,312] as well as accelerator-based techniques. By combining the high peak power femtosecond lasers with relativistic electron beams, tunable x-rays with improved flux are realized by exploiting 90-degrees Thomson scattering,[313,314] using lasers as energy modulators in electron storage rings,[315,316] and by the integration of ultrafast detectors into synchrotron beamlines.[317] Accelerator-based sources provided multiple-order-of-magnitude improvements in the x-ray flux and enabled the demonstration of x-rays from free-electron laser (XFEL) operation. The XFEL process can be understood similarly to an optical laser but with relativistic electrons as the gain medium. It relies on the interaction between the radiation field of an electron passing through an undulator and the electron itself. The core of FEL operation is to amplify the spontaneous radiation in a feedback process by re-interacting it with the electrons such that electron bunching and x-ray growth self-consistently drive each other, leading to an exponential increase in overall radiation intensity.[279,318] Femtosecond x-ray pulses scattered, transmitted, or nonlinearly-converted by the sample can be analyzed in multimodal perspectives under different detector configurations, e.g., momentum transfer, spectrum, coherence, etc., which contains the fingerprint of the ultrafast dynamics triggered by one or more ultrafast pump pulses.

Therefore, in these schemes, femtosecond x-rays provide another route complementary to the ultrafast electron microscope techniques that can access the dynamics of structural, electronic, and magnetic degrees of freedom, which is the key to understanding the underlying non-equilibrium physics behind exotic phenomena in quantum materials. Although both XFEL and UEM/UED methodologies can assist in tackling challenges in quantum science that entail atomic-level views and are sometimes interchangeable, it is important to emphasize their differences and further their potential to be combined to yield a more comprehensive understanding.

One main difference between these techniques is that the scattering cross-section for x-rays is about six orders of magnitude weaker than the cross-section for electrons. Therefore, multiple scattering events play a less important role for x-ray scattering, and the kinematic approximation still applies in most cases. This dramatically simplifies the quantitative analysis of the x-ray diffraction intensity. Although several electron microscope designs have leveraged various workarounds,[289,319] this dynamic effect still poses challenges in electron scattering analysis such as deviations from Debye-Waller behavior,[320] the appearance of symmetry-forbidden Bragg peaks, and a seeming violation of Friedel's Law even in the monolayer limit.[321] This is also a part of the reason electron transparent samples are required for electron microscopy studies. In contrast, x-ray scattering allows macroscopically thick samples to be examined with minimal sample preparation complications. For example, exotic quantum properties in oxide perovskites change at different thicknesses.[322–324] In those experiments, x-rays have been applied to unveil dynamics of various classes of quantum materials in their bulk form. Quantum orders including charge,[325,326] ferroic,[327,328] lattice,[329,330] spin,[331–333] and orbital[329,330,334,335] orders have been scrutinized by x-ray in materials like complex oxides, correlated materials, and superconductors, among others.

Another significant difference between XFEL and UEM is the space charge effect from the collective Coulomb repulsion of free charges in space or on a surface, which remains a persistent challenge in UEM and UED single-shot mode. This detrimental effect limits temporal and momentum resolution for UEM and UED. Attosecond x-ray pulses[336,337] from FELs can now be routinely achieved[336,337] while overcoming the fs temporal resolution limit remains an ongoing target for UEM.[304,319] Thus, x-rays can complement ultrafast electron measurements to capture the macroscale few-femtosecond dynamics in quantum materials such as phase transitions in correlated electron systems.[338] Also, coherent x-ray beams from XFELs greatly surpass ultrafast electron microscopes in terms of momentum resolution and have enabled inelastic/diffuse scattering-type probes[325,339] and capture of lattice orders with long length-scale periodicities [340] via reciprocal space mapping techniques. Meanwhile, the MeV electron energies in UED/UEM allow a flatter Ewald's sphere and simultaneous monitoring of multiple Bragg peaks. This is advantageous as peaks at different orders carry distinct information about material characteristics and can enable a tomographic reconstruction of the unit cell. Therefore, we anticipate comprehensive dynamics studies with x-rays and electron microscopy complementing each other, covering broad, spatial, momentum, temporal, and energy regimes.

There are growing examples of combining results from x-ray and ultrafast electron experiments to tackle ongoing scientific challenges. For example, the photoinduced insulator-to-metal transition in the correlated electron material $VO_2$ was studied with UED,[293,295] and follow-up x-ray diffuse scattering measurements yielded a detailed order-disorder transformation mechanism.[341] For two-dimensional materials, femtosecond x-rays can provide out-of-plane dynamics [342] and complement the typical in-plane dynamics obtained with UED.[343,344] Electron beams undergo a dramatic streaking effect under terahertz or long-wavelength infrared pulse excitation. This poses a significant drawback in unveiling structural dynamics close to or within the excitation pulse and can be tackled with x-rays.[324,345,346] The continuous improvement in longitudinal and transverse coherence in X-rays from XFELs has enabled multiple tools, such as x-ray coherent diffractive imaging[347–349] and x-ray photon correlation spectroscopy,[350,351] that can access mesoscale domain or aging dynamics in systems close to a phase transition. Also, the fast-developing attosecond x-ray sources can achieve unprecedented temporal resolution[336,337,352] while also enabling spectroscopic probes with elemental specificity. Exploring these novel tools in a time-dependent fashion can provide a new view of nonequilibrium dynamics in quantum materials that cannot be readily provided by an ultrafast electron microscope.

Besides x-rays, ultrafast optical spectroscopy contains dynamic and often non-structural information complementing or supporting UEM data. Thorough summaries of ultrafast optical spectroscopies in quantum materials can be found in other reviews.[353,354] Widely used observables are linear probes, including transmission, reflection, emission, and current, which carry information directly related to quantum property changes upon structural distortion. For example, DC or terahertz conductivity inherits fingerprints of superconductivity,[355,356] metallicity,[357–359] ferroelectricity,[323,360] magnetism,[361,362], etc.; the visible dielectric function bears signatures of plasmons,[363,364] excitons,[365,366] and other elementary excitations. With the structural dynamics obtained from UEM, a more comprehensive perspective of the material response and far-from-equilibrium excursion can be acquired. Nonlinear optical observables have also been adopted to probe symmetry-related responses upon excitation in quantum materials, notably second harmonic generation,[298,323,367,368] Kerr effect,[323,369,370] and time-resolved Raman scattering techniques.[371,372] A representative example is presented in work by Sie et al., where light-induced large-amplitude shear oscillations in a Weyl semimetal $WTe_2$ were observed using both UED (Figure 9e-g) and a second harmonic generation probe (Figure 9h).[298] Figure 9e shows that the quantitative calibration of the shear amplitude and symmetry change demonstrated by the SHG probe corroboratively indicates an ultrafast symmetry switch by light, in turn modulating the unique topological properties of $WTe_2$. Their successful marriage in quantum material research shows that these nonlinear optical methods can potentially support next generation electron microscopy explorations.

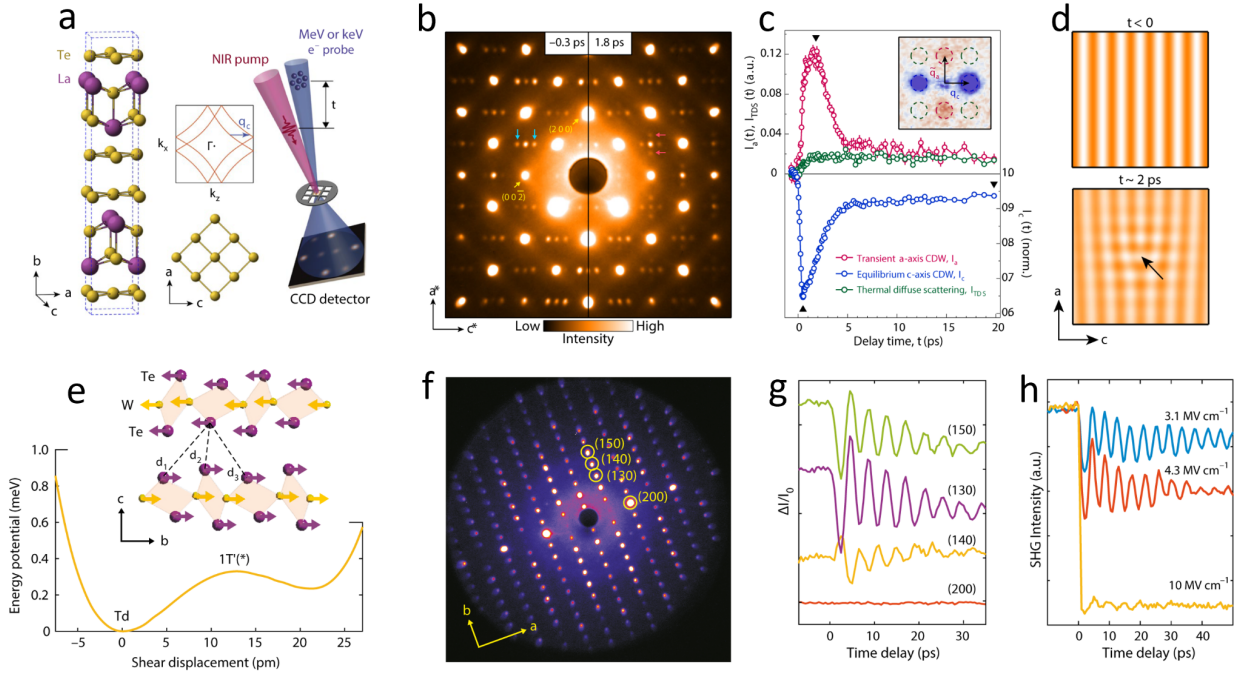

Figure 9. (a) 800-nm pump UED probe on LaTe$_3$. (b) The electron diffraction patterns before (left) and 1.8 ps after (right) photoexcitation with 800-nm pulse excitation, blue and red arrows indicate the equilibrium CDW peaks along the *c* axis and the light-induced CDW peaks along the *a*-axis, respectively. (c) Time evolution of integrated intensities of the equilibrium *c*-axis CDW peak ($I_c$), the transient *a*-axis CDW peak ($I_a$) and thermal diffuse scattering ($I_{TDS}$). Integration areas are marked in the inset. (d) Schematics of charge density waves in real space before (top) and ~2 ps after (bottom) photoexcitation seeded by topological defects. Adapted from ref.[290]. Copyright 2020 Nature publishing group. (e) The interlayer shear displacements in Weyl semimetal WTe$_2$ and associated energy potential landscape. (f) Static electron diffraction pattern of WTe$_2$. (g) Visualization of the light-induced shear oscillations with UED. (h) Second harmonic generation probe of the symmetry switch. Adapted from ref.[298]. Copyright 2019 Nature publishing group.

## 7. Deriving 3D material structure, including quantum defects using atomic electron tomography (AET)

TEM imaging and spectroscopic techniques usually provide structural, chemical, and electronic information through projected 2D images. A thorough understanding of structural defects and their impact on macroscopic physical properties requires the ability to accurately measure 3D atomic coordinates. Such an ability poses the potential for the study of defects at the single atom level.

Electron tomography, a technique historically used for imaging 3D structures of materials, utilizes 2D projections of an object over a wide range of tilt angles to reconstruct a 3D structure.[373–375] The resolution of electron tomography reconstruction is determined by the range and number of tilt angles, the electron dose, and the resolution of 2D projected images. This technique has been developed for decades and has enabled imaging 3D structures of nanostructures with nm resolution. The use of aberration-corrected S/TEM and direct electron detectors, in combination with advanced reconstruction algorithms, has led to the recent development of atomic electron tomography (AET) that enables 3D imaging of atomic structures, crystal defects, and mixed-order systems with sub-Å resolution.[376]

AET has been used to study crystal defects such as grain boundaries, dislocations, stacking faults, point defects, and strain tensors in 2D materials and magnetic materials with unprecedented detail.[377–381] In a recent study, Miao and collaborators used the method to measure 3D atomic displacements in 2D $MoS_2$ thin films doped with individual Re atoms with a precision down to a few pm.[382] Measurement of local 3D structure of the $MoS_2$ lattice around the Re dopants allows the determination of the local strain tensor and local deviations of bond distances/angles around Re atoms. The results show that Re dopants generate a larger out-of-plane strain than the in-plane strain, providing insights into the mechanism of the dopant-induced 2H to 1T phase transition in the TMDCs.

The measured atomic coordinates from AET can be used as an input into ab initio calculations to reveal functional properties of materials down to the single-atom level. In the above mentioned study of Re-doped $MoS_2$, the obtained 3D structural data has been used as an input in density functional theory (DFT) simulations to calculate local band diagrams. The calculation has demonstrated that the presence of dopants causes a localized metal-to-insulator transition at the atomic scale. Yang et al. has applied AET to precisely determine the 3D coordinates and elemental arrangement of a magnetic FePt nanoparticle.[380] DFT calculations based on the measured 3D atomic map have revealed the local variation of spin and orbital moment at the atomic scale. AET, in combination with DFT, is anticipated to provide key insight into the 3D atomic structure–property relationship in both equilibrium and metastable quantum materials.

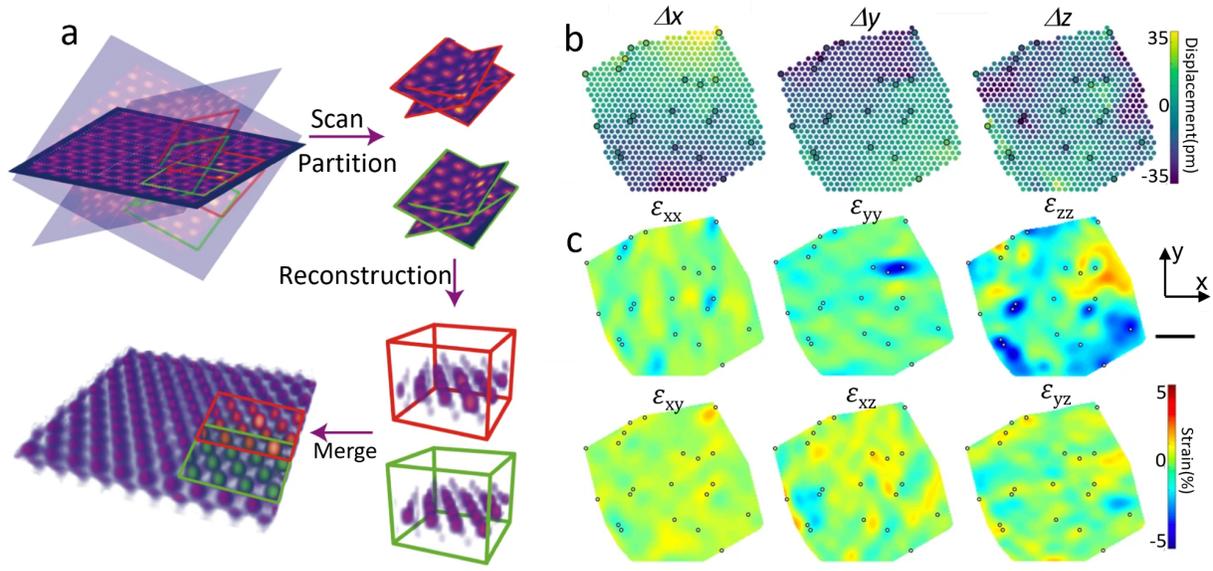

Figure 10. (a) Schematic of the scanning atomic electron tomography. A limited number of projections at different tilt angles are acquired from a 2D material or heterostructure. A 3D window is then chosen and scanned along the $x$ and $y$ axes, maintaining a finite overlap between two adjacent windows. At each scanning position, corresponding regions in the projections are identified for each window and all the projections are partitioned into a series of image stacks. The 3D windows are reconstructed from the image stacks and stitched together to form a full reconstruction of the atomic structure. (b) Atomic displacements of Mo (dots) and Re (circled dots) atoms along the $x$, $y$ and $z$ axes, respectively. (c) Six components of the strain tensor in the Mo/Re layer, where the Re dopants (circles) induce local strains in the $\varepsilon_{xx}$, $\varepsilon_{yy}$ and $\varepsilon_{zz}$ maps. Reproduced with permission from ref.[382,383]. Copyright 2020 Nature publishing group.

## 8. Summary and Outlook

The realization of genuine quantum effects in solid-state materials hinges on the fine control of electronic coherences in real-, time-, energy- and momentum-space, calling for materials characterization techniques that probe the structure and dynamics down to the intrinsic limits of *single atoms*. In this review, we outlined recent advances in electron microscopy, coupled optical/electron microscopy, time-resolved electron microscopy and complementary techniques including UED and XFEL spectroscopies. These techniques provide excellent tools for studying quantum materials and excitations spanning lattice, spin, charge, orbital, electronic, and topological orders. By implementing such techniques at sub-atomic length scales, femtosecond temporal resolution, and single-digit meV energy resolution, deep understanding of fundamental quantum excitations have been gained. For example, researchers have been able to measure the local 3D structure and heterogeneities with picometer precision, directly visualizing the formation of the topological quasiparticles and transient states, directly image Bloch modes and spin crossovers, map real space charge-density waves with atomic resolution, quantitatively measure interatomic electric field.

Maintaining quantum phase-relationships between electrons in materials requires precise control over dephasing, naturally achieved at low temperatures. Combining extremely low temperatures with high stability and thereby enabling atomic-resolution capabilities at ultralow T could enable the next boom in the world of electron microscopy. Being able to take advantage of full EM capabilities at cryogenic temperatures (ultra-lowT) requires significant instrumental advancements to overcome stage-stability issues in Cryo-EM.[59,60] Minimizing drift and vibration combined with the ability to reach ulta-low temperatures ideally down to 1K will significantly benefit the investigation of the next generation of quantum materials. Such advancements will open new avenues for exploring exotic properties of quantum materials at relevant length scales and at extremely low temperatures (~93K down to a few Kelvin). Once these challenges have been overcome, one can explore the microscopic origin of electron-lattice interactions, spatially resolve charge and orbital ordering, and directly visualize a myriad range of exotic quantum and electronic phenomena ranging from superconductivity to metal-insulator transitions.[60,191,193,384–388] Ultimately, Cryo-S/TEM will provide a roadmap on how to modulate and alter the structure of quantum materials in order to tune and improve the functional properties of such systems.

We anticipate the development of advanced multi-modal characterization tools including combined electron, optical and X-ray based spectroscopy and imaging techniques will drive the development of new discoveries in quantum materials. Such complementary modalities will not only overcome the limitations of each technique but also combine and incorporate the capabilities of each platform. The deployment of a multi-modal characterization will enable extracting multi-scale and multi-dimensional information from emerging quantum materials in their native and working condition (ultra-low temperature). EM will continue as a prominent toolkit to learn the strongly correlated material, to expedite many discoveries on the atomic scale as well as to identify the root cause behind many discoveries.The coming paradigm shift in EM will herald a new era in quantum metrology- spanning materials synthesis, sensors, and systems.

**Acknowledgements**

P.M. acknowledge the support from the Stanford Bio-X seed grant under award no. 1248231-100-WXDCW and Stanford Molecular Imaging Scholars Program (SMIS) under award no. NIH T32 CA118681. J.S., H.U., A.M.L. and J.A.D. acknowledge support from the Photonics at Thermodynamic Limits Energy Frontier Research Center, funded by the U.S. Department of Energy, Office of Science, Office of Basic Energy Sciences, under award no. DE-SC0019140. J.S. and A.M.L. acknowledge support by the Department of Energy, Office of Science, Basic Energy Sciences, Materials Sciences and Engineering Division, under Contract DE-AC02-76SF00515.  In addition, the authors acknowledge support from the Q-next grant under award no. DE-AC02-76SF00515. We thank Kazu Suenaga and Ryosuke Senga for providing figures on the phonon band structures of graphene and hBN obtained via EELS.

## Author Information

## Corresponding Author

- **Parivash Moradifar -** *Department of Materials Science and Engineering, Stanford University School of Engineering, Stanford, California 94305, United States*; https://orcid.org/0000-0001-7445-4628; Email: pmoradi@stanford.edu

- **Jennifer A. Dionne -** *Department of Materials Science and Engineering, Stanford University School of Engineering, Stanford, California 94305, United States*; *Department of Radiology, Stanford University School of Medicine, Stanford, California 94305, United States*; https://orcid.org/0000-0001-5287-4357; Email: jdionne@stanford.edu

## Authors

**Parivash Moradifar-***Department of Materials Science and Engineering, Stanford University, Stanford, California 94305, USA*

**Yin Liu-***Department of Materials Science and Engineering, Stanford University, Stanford, California 94305, USA; Department of Materials Science and Engineering, North Carolina State University, Raleigh, North Carolina 27695, USA*

**Jiaojian Shi-***Department of Materials Science and Engineering, Stanford University, Stanford, California 94305, USA; SLAC National Accelerator Laboratory, 2575 Sand Hill Road MS69, Menlo Park, California 94025, USA*

**Matti Lawton Siukola Thurston-***Department of Materials Science and Engineering, Stanford University, Stanford, California 94305, USA*

**Hendrik Utzat-***Department of Materials Science and Engineering, Stanford University, Stanford, California 94305, USA; Department of Chemistry, University of California Berkeley, Berkeley, California 94720, USA*

**Tim B. van Driel-***Linac Coherent Light Source, SLAC National Accelerator Laboratory, 2575 Sand Hill Road, Menlo Park, California 94025, USA*

**Aaron M. Lindenberg-***Department of Materials Science and Engineering, Stanford University, Stanford, California 94305, USA; SLAC National Accelerator Laboratory, 2575 Sand Hill Road MS69, Menlo Park, California 94025, USA*

**Jennifer A. Dionne-***Department of Materials Science and Engineering, Stanford University, Stanford, California 94305, USA; Department of Radiology, Stanford University, Stanford, California 94305, USA*

**Authors contribution**

P.M. and J.A.D. defined the scope of the manuscript. P.M., J.S., Y.L. refined the scope in particular sections and areas of expertise. P.M. drafted the initial manuscript with input from J.S., Y.L., H.U., M.L.S.T. J.A.D., A.M.L. and T.Bv.D. supervised the work. All authors read and approved the paper.

**Biographies**

Parivash Moradifar is a postdoctoral researcher and SMIS fellow in Materials Science and Engineering at Stanford University. Parivash received her Ph.D. in Materials Science and Engineering from Pennsylvania State University in 2020 where she used TEM as a microscale laboratory to research the localized plasmonic behavior of low-dimensional topological insulators and extended nano-assemblies to identify the impact of physical and chemical modifications. She is passionate to push electron microscopy beyond imaging and bridge electron microscopy, X-ray science and ultrafast spectroscopies to develop new nanomaterials as well as to understand the elementary excitations under dynamic conditions.

Yin Liu is an assistant professor in the department of Materials Science and Engineering at North Carolina state university. He received his Ph.D. in Materials Science and Engineering from UC Berkeley in 2019. He conducted his postdoctoral work with Prof. Jennifer Dionne in the Department of Materials Science and Engineering at Stanford University. His current research is focused on optoelectronic low-dimensional materials, with a particular interest in combining TEM imaging and spectroscopy with optical spectroscopy to interrogate optical properties of materials on the nanoscale.

Jiaojian Shi is a postdoctoral researcher in the Department of Material Science and Engineering at Stanford University and the SLAC National Accelerator Laboratory. He received his Ph.D. in physical chemistry from the Massachusetts Institute of Technology in 2021, where he studied strong-field phenomena at terahertz frequencies in single nanocrystals and low-dimensional quantum materials. His current interest is probing the ultrafast dynamics of single quantum emitters in two-dimensional materials with structural-sensitive spectroscopies.

Matti Thurston is a graduate researcher in the Department of Materials Science and Engineering at Stanford University currently working with Prof. Jennifer Dionne. He received his B.S in materials science and engineering from Cornell University in 2018, during which time he studied the charge transport properties of thin film cuprate superconductors in Schlom lab, followed by work as a project manager at Apple Inc. His current work focuses on nanophotonics and related semiconductor materials for application in next generation optics, quantum information systems, and computers.

Hendrik Utzat is an Assistant Professor of Chemistry at the College of Chemistry at UC Berkeley. He received his Ph.D. in physical chemistry from the Massachusetts Institute of Technology (MIT) in 2019. His work focused on the development of optical single-emitter spectroscopy and the elucidation of optical coherences in single colloidal quantum dots and quantum defects in two-dimensional materials. Hendrik conducted postdoctoral work in the

Department of Materials Science and Engineering at Stanford University studying the structure-property relationships of silicon-vacancy centers in diamond using optically-coupled electron microscopy. Hendrik joined the UC Berkeley College of Chemistry in July 2022.

Tim B. van Driel is a staff scientist at SLAC National Accelerator Laboratory. He received his PhD in Physics from the Technical University of Denmark in 2014. His work focuses on ultrafast X-ray science, in particular, using a combination of scattering and spectroscopic techniques to study solution phase chemistry as well as develop and improve on existing X-ray free electron laser capabilities .

Aaron Lindenberg is a Professor of Materials Science and Engineering and of Photon Science at Stanford University and the SLAC National Accelerator Laboratory. He received his Ph.D. in physics from the University of California, Berkeley, in 2001. His work focuses on understanding the ultrafast properties of materials.

Jennifer Dionne is the Senior Associate Vice Provost of Research Platforms/Shared Facilities at Stanford, and an associate professor of Materials Science and Engineering and, by courtesy, of Radiology. She is also a Chan Zuckerberg Biohub Investigator and an Associate Editor of Nano Letters. Jen received her Ph.D. in Applied Physics at the California Institute of Technology in 2009, and her postdoctoral training in Chemistry at Berkeley.  As a pioneer of nanophotonics, she is passionate about developing methods to observe and control chemical and biological processes as they unfold with nanometer scale resolution, emphasizing critical challenges in global health and sustainability. Jen is passionate about translating scientific inventions to commercial innovations, and is co-founder of a company enabling life-speed reads of biological bits.

**Notes**

The authors declare no competing financial interest.